%% file: main_latest.tex
\documentclass[]{aastex631}

\input{mypreset}

\acsetup{patch / longtable = false }

\received{\today}
\revised{MM DD, YYYY}
\accepted{MM DD, YYYY}

\submitjournal{ApJ}

\shorttitle{
\j1641
}
\shortauthors{Tsuji et al.}
\graphicspath{{./}{figures/}}

\begin{document}

\title{
Search for synchrotron emission from secondary electrons of proton-proton interaction
in Galactic PeVatron candidate HESS J1641$-$463 
}



\correspondingauthor{Naomi Tsuji}
\email{ntsuji@kanagawa-u.ac.jp}

\author[0000-0002-0786-7307]{Naomi Tsuji}
\affiliation{Faculty of Science, Kanagawa University, 3-27-1 Rokukakubashi, Kanagawa-ku, Yokohama-shi, Kanagawa 221-8686, Japan}
\affiliation{Interdisciplinary Theoretical \& Mathematical Science Program (iTHEMS), RIKEN, 2-1 Hirosawa, Wako, Saitama 351-0198, Japan}

\author{Takaaki Tanaka}
\affiliation{Konan University, Department of Physics, 8-9-1 Okamoto, Higashinada, Kobe, Hyogo, Japan, 658- 8501}

\author{Samar Safi-Harb}
\affiliation{Department of Physics and Astronomy, University of Manitoba, Winnipeg, MB R3T 2N2, Canada}

\author{Felix Aharonian}
\affiliation{Dublin Institute for Advanced Studies, School of Cosmic Physics, 31 Fitzwilliam Place, Dublin 2, Ireland}
\affiliation{Max-Planck-Institut f\"{u}r Kernphysik, Saupfercheckweg 1, 69117 Heidelberg, Germany}

\author{Sabrina Casanova}
\affiliation{Instytut Fizyki J\c{a}drowej PAN, ul. Radzikowskiego 152, 31-342 Krak{\'o}w, Poland}

\author[0000-0001-5953-0100]{Roland Kothes}
\affiliation{Dominion Radio Astrophysical Observatory, Herzberg Astronomy and Astrophysics Research Centre, National Research Council Canada, P.O. Box 248, Penticton, British Columbia
V2A 6J9, Canada }

\author{Emmanuel Moulin}
\affiliation{IRFU, CEA, Universit\'e Paris-Saclay, F-91191 Gif-sur-Yvette, France}

\author{Hiroyuki Uchida}
\affiliation{Department of Physics, Kyoto University, Kitashirakawa Oiwake-cho, Sakyo, Kyoto 606-8502, Japan}

\author{Yasunobu Uchiyama}
\affiliation{Department of Physics, Rikkyo University, 3-34-1 Nishi Ikebukuro, Toshima-ku, Tokyo 171-8501, Japan}



\begin{abstract}
\j1641 is an unidentified gamma-ray source with a hard TeV gamma-ray spectrum, and thus it has been proposed to be a possible candidate for \ac{cr} accelerators up to PeV energies (a PeVatron candidate). 
The source spatially coincides with the radio \ac{snr} G338.5$+$0.1, but has not yet been fully explored in the X-ray band. 
We analyzed newly taken \nustar\ data, pointing at \j1641, with 82 ks effective exposure time.
There is no apparent X-ray counterpart of \j1641, while nearby stellar cluster, Mercer~81, and stray-light X-rays are detected.
Combined with the archival \chandra\ data, partially covering the source, we derived an upper limit of 
$\sim 6\times 10^{-13}$~\flux\ in 2--10 keV 
($\sim 3\times 10^{-13}$~\flux\ in 10--20 keV).
If the gamma-ray emission is originated from decay of \pizero\ mesons produced in interactions between \ac{cr} protons and ambient materials, secondary electrons in the proton-proton interactions can potentially emit synchrotron photons in the X-ray band, which can be tested by our X-ray observations.
Although the obtained X-ray upper limits cannot place a constraint on the primary proton spectrum, it will be possible with a future hard X-ray mission.


\end{abstract}


\section{Introduction} 
\label{sec:intro}
\acresetall

The origin of high-energy \acp{cr} remains one of the most important topics in high-energy and astroparticle physics.
CRs up to the knee ($\sim$3 PeV) have been widely believed to be produced in \acp{snr} in the Galaxy through the diffusive shock acceleration mechanism \citep[e.g.,][]{vink_supernova_2011}.
One of the issues in this SNR paradigm is that most of the \acp{snr} detected in the gamma-ray range have spectra with exponential cutoff of at most a few tens of TeV, which failed to give a decisive answer to whether SNR shocks are capable of accelerating particles up to the knee ---PeVatrons \citep[e.g.,][]{Funk2015}.
However, a remarkable discovery was made by the High Energy Stereoscopic System (\hess) collaboration: 
Relatively hard spectra without any sign of cutoff were obtained from the Galactic center \citep{HESS2016_gc} and the unidentified source \j1641\ \citep{HESS2014_hessj1641}, 
making them as the first candidates for PeVatron sources.
Recently, (sub-) PeV gamma-ray astronomy has begun by extensive air shower arrays (EASA), such as HAWC, Tibet AS$\gamma$, and LHAASO.
These observations are accelerating the study of PeVatrons.

In this paper, we would like to revive the interest in the primary PeVatron candidate, \j1641\footnote{Note that \j1641\ is located at a position where LHAASO and other EASAs on the north hemisphere cannot observe.} \citep{HESS2014_hessj1641}.
Since it is close to another gamma-ray bright source HESS~J1640$-$465 \citep{abramowski_hess_2014}, the confusion had been preventing the detection of \j1641\ until recently. 
Thanks to the improved \hess\ point spread function, 
\j1641\ was discovered in the tail of HESS~J1640$-$465 with a slight spatial extension up to $\sim$3\arcmin\ \citep{HESS2014_hessj1641,HESS2018_HGPS}.
\j1641\ spatially coincides with the \ac{snr} G338.5$+$0.1, which has roughly a circular morphology with a diameter of 5\arcmin--9\arcmin\ in the radio band \citep{whiteoak_most_1996}, a distance of $\approx$11 kpc \citep{kothes_distance_2007}, and an estimated age of 1.1--17 kyr \citep{HESS2014_hessj1641}.
\j1641\ has one of the hardest gamma-ray spectra with
$\Gamma = 2.07 \pm 0.11_{\rm stat} \pm 0.20_{\rm sys}$ in 0.64--100 TeV without any indication of a cutoff, implying that the cutoff energy of accelerated particles is higher than 100 TeV at 99\% C.L. \citep{HESS2014_hessj1641}.
Thus, \j1641\ was proposed to be a primary candidate for a PeVatron.

The origin of gamma-ray emission in the \j1641\ and HESS~J1640$-$465 system remains unsettled.
In the GeV gamma-ray band,
there exist counterparts of \j1641\ that has a point-like morphology and log-parabola spectrum and of HESS~J1640$-$465 that is characterized by $\sim$3\farcm2 spatial extension and power-law spectrum with $\Gamma =1.7$ \citep{mares_constraining_2021}.
As discussed by \cite{mares_constraining_2021},
there are a couple of scenarios for this complicated system, such as (1) hadronic gamma-ray radiation (\ac{snr} or illuminated molecular cloud), (2) pulsar, or (3) PWN for the emission of \j1641. 
None of them, however, are conclusive for \j1641, while HESS~J1640$-$465 is likely originated from a PWN \cite[e.g.,][]{abdelmaguid_broadband_2023}.

The spectrum without a cutoff generally prefers a hadronic gamma-ray production over leptonic emission (i.e., \ac{ic} scattering of the ambient photons by accelereated electrons) because the Klein-Nishina effect significantly reduces the gamma-ray flux and inevitably makes a spectral cutoff at $\gtrsim$10 TeV.
The hadronic gamma-ray emission is attributed to decay of neutral pions produced in interactions of \ac{cr} protons with ambient protons.
The same proton-proton interactions produce charged pions, which decay into muons and electrons/positrons in turn.
These electrons and positrons are often referred to as \textit{secondary electrons}.
If the energy of the parent proton is sufficiently high, the secondary electrons are also energetic enough to emit X-ray photons via synchrotron radiation, which can be potentially measured with X-ray observations.
Although there are some dedicated studies to predict the secondary synchrotron component \citep[e.g.,][]{aharonian_very_2004,aharonian_gamma_2013, huang_secondary-electron_2020},
the emission has not been detected to date.
In this paper, we aim at searching for the secondary synchrotron emission in the X-ray data of \j1641\ and
investigating how the synchrotron radiation from the secondary electrons depends on physical parameters.
If detected, we show that the X-ray observation would be a new diagnostic tool to determine the maximum energy of accelerated protons,
while only lower limit ($>$100 TeV) is obtained by the current gamma-ray observation.

We present X-ray observations of \j1641\ in \secref{sec:xray}.
\secref{sec:modeling} describes modeling of the broadband \ac{sed}, including calculation of the synchrotron radiation from secondary electrons.
The discussion and conclusion are presented in Sections \ref{sec:discussion} and \ref{sec:conclusion}, respectively.

\section{X-ray analysis} \label{sec:xray}

This section describes X-ray observations and analysis of the gamma-ray source \j1641.
In the soft X-ray band ($\lesssim 10$ keV), on-axis observations pointing at this source have not been performed.
There are archival observations with \chandra\ (Obs IDs 11008 and 12508) and \xmm\ (Obs ID 0302560201), which partially cover the gamma-ray extent of \j1641.
We made use of these \chandra\ data in this paper.\footnote{We do not use the \xmm\ observation because of its large ($\sim 14$\arcmin) off-axis pointing.}
In addition to the soft X-ray data, we also used newly retrieved hard X-ray data by \nustar\ (Obs ID 40401004002).
\tabref{tab:dataset} summarizes information on these X-ray data analyzed in this paper.

%
The \chandra\ data were processed using CALDB version 4.9.6 in CIAO version 4.14.
The \nustar\ data were calibrated and screened by using nupipeline of NuSTAR Data Analysis Software (NuSTARDAS version 2.1.1 with CALDB version 20220118) included in HEAsoft version 6.29. 
We used the strictest mode (i.e., SAAMODE$=$STRICT and TENTACLE$=$YES cut) for screening the \nustar\ data. 
We present the analysis and results of X-ray images and spectra in Sections \ref{sec:image} and \ref{sec:spectrum}, respectively.

\begin{deluxetable*}{ cccccccc }
\tablecaption{
Dataset of X-ray observations
\label{tab:dataset}
}
\tablewidth{0pt}
\tablehead{
\colhead{Satellite} &
\colhead{Target name} & \colhead{Obs ID} & \colhead{Effective exposure} & \colhead{Date} & \colhead{RA}  & \colhead{Dec}  & \colhead{Roll angle}  \\ 
& & & \colhead{(ks)} & & \colhead{(deg)} & \colhead{(deg)} & \colhead{(deg)}
}
\startdata
\chandra & GLIMPSE~81 &  11008 &  39.6 &  2010-06-19 &  250.1 &  $-$46.4 &  321.7 \\    
\chandra & Norma Region &  12508 &  18.5 &  2011-06-06 &  250.2 &  $-$46.5 &  342.2 \\ 
\nustar &  \j1641 &  40401004002 &  81.6 &  2018-10-11 &  250.2 &  $-$46.3 &  357.8 \\    
\enddata
\end{deluxetable*}

\subsection{Image} \label{sec:image}

We used {\tt fluximage} to produce \chandra\ flux images.
For \nustar\ images, we generated count maps and exposure maps by Xselect and {\tt nuexpomap}, respectively, and then produced flux images by dividing the count maps by the exposure maps.  
Background was not taken into account for the X-ray flux images.

\begin{figure}[ht!]
\gridline{
    \fig{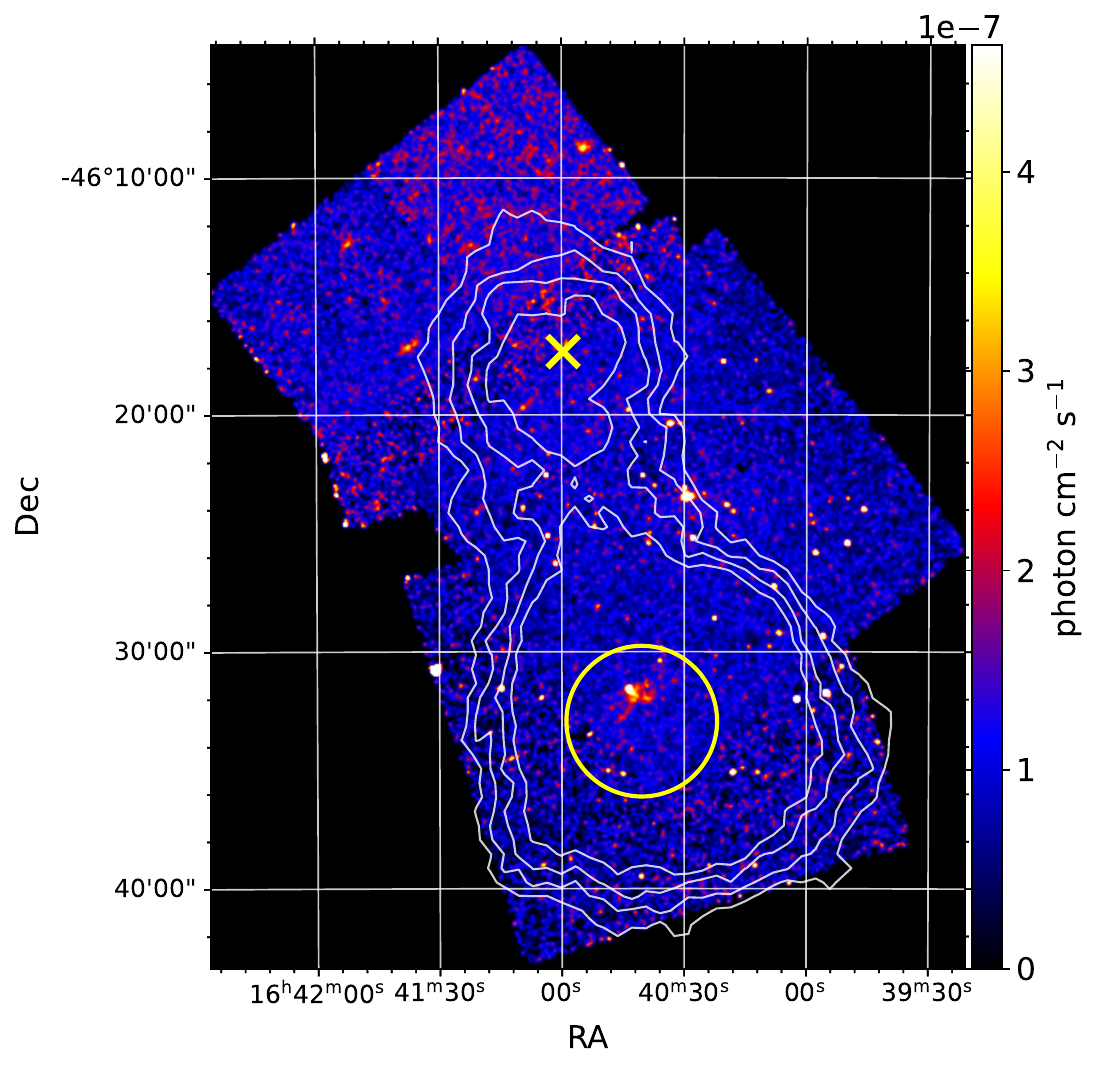}{0.33\textwidth}{(a) \chandra\ 0.5--7 keV}
    \fig{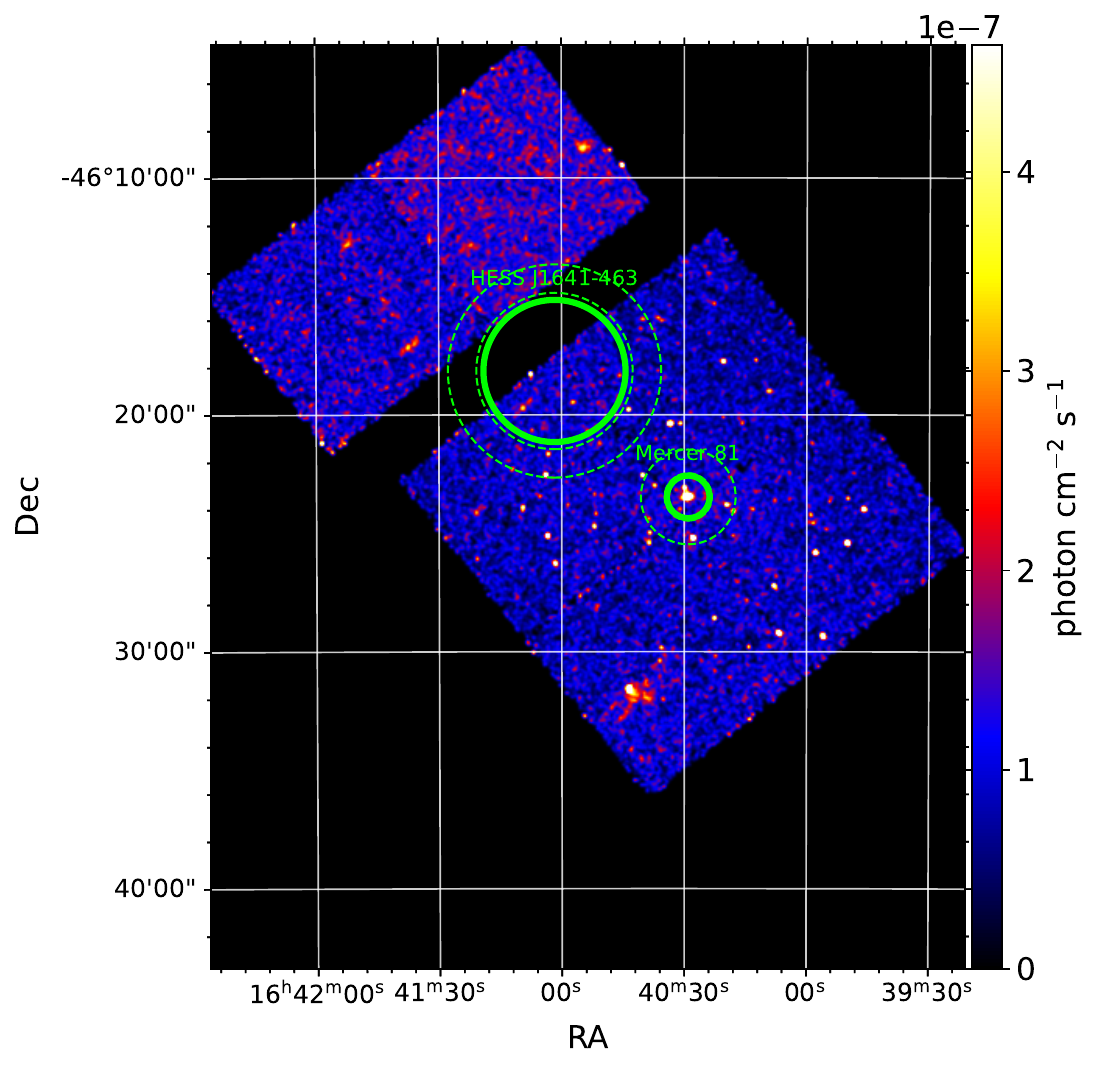}{0.33\textwidth}{(b) \chandra\ (ObsIDs 11008) 0.5--7 keV}
    \fig{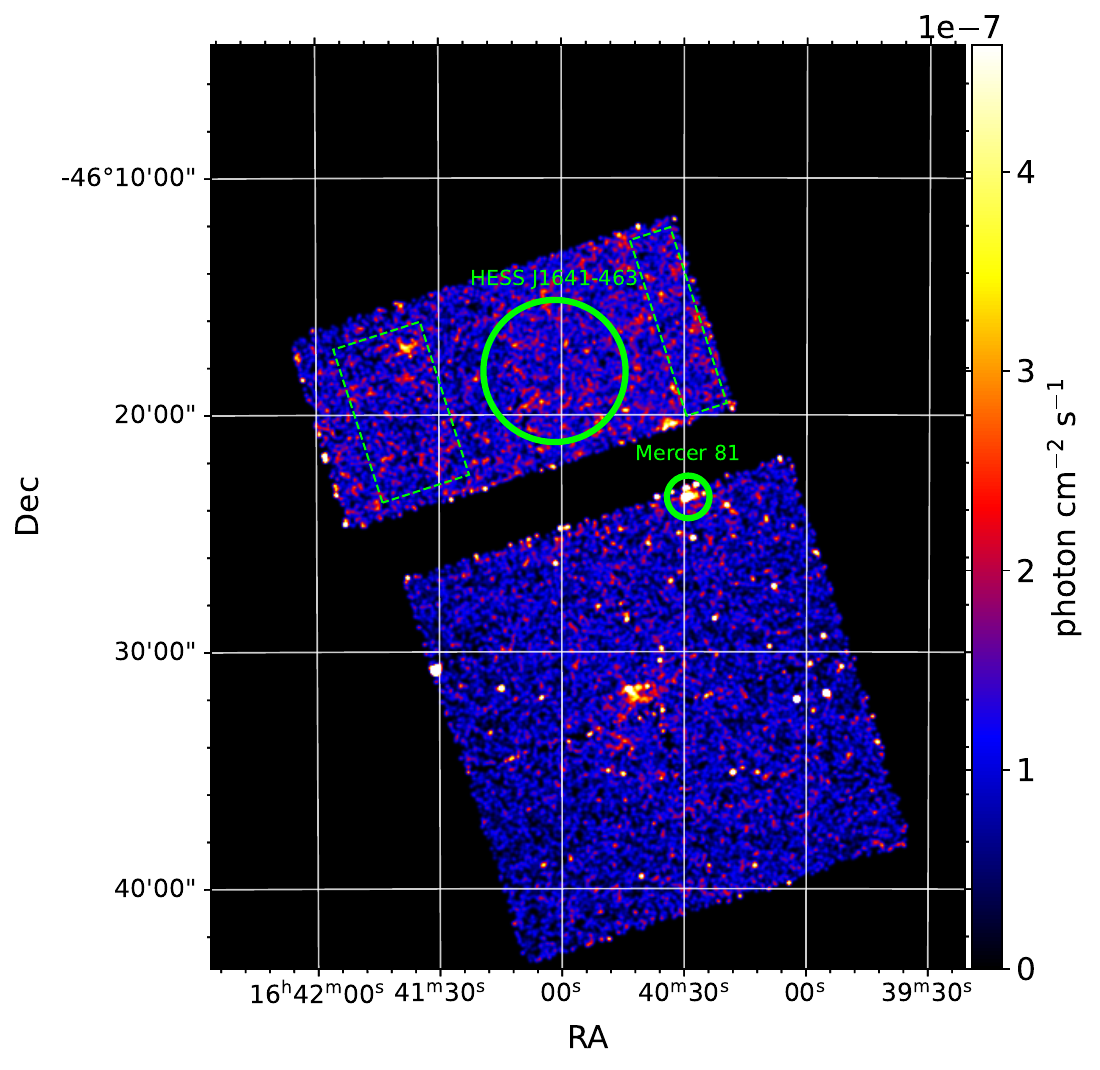}{0.33\textwidth}{(c) \chandra\ (ObsIDs 12508) 0.5--7 keV}
          }
\caption{
Flux images of \j1641\ in 0.5--7 keV with \chandra\ ObsIDs 11008 (b), 12508 (c), and merged (a). 
(a):
The contour indicates the significance map of \hess\ \citep{HESS2014_hessj1641}.
The yellow cross and circle show  positions of \j1641\ and HESS~J1640$-$465 obtained by the \lat\ data \citep{mares_constraining_2021}, respectively.
(b) and (c):
The large green circle shows the TeV gamma-ray extent of \j1641\ with its radius of 3\arcmin.
Source and background regions for spectral analysis are indicated by green solid and green dashed lines, respectively.
\label{fig:image_chandra} }
\end{figure}

\begin{figure}[ht!]
\gridline{
    \fig{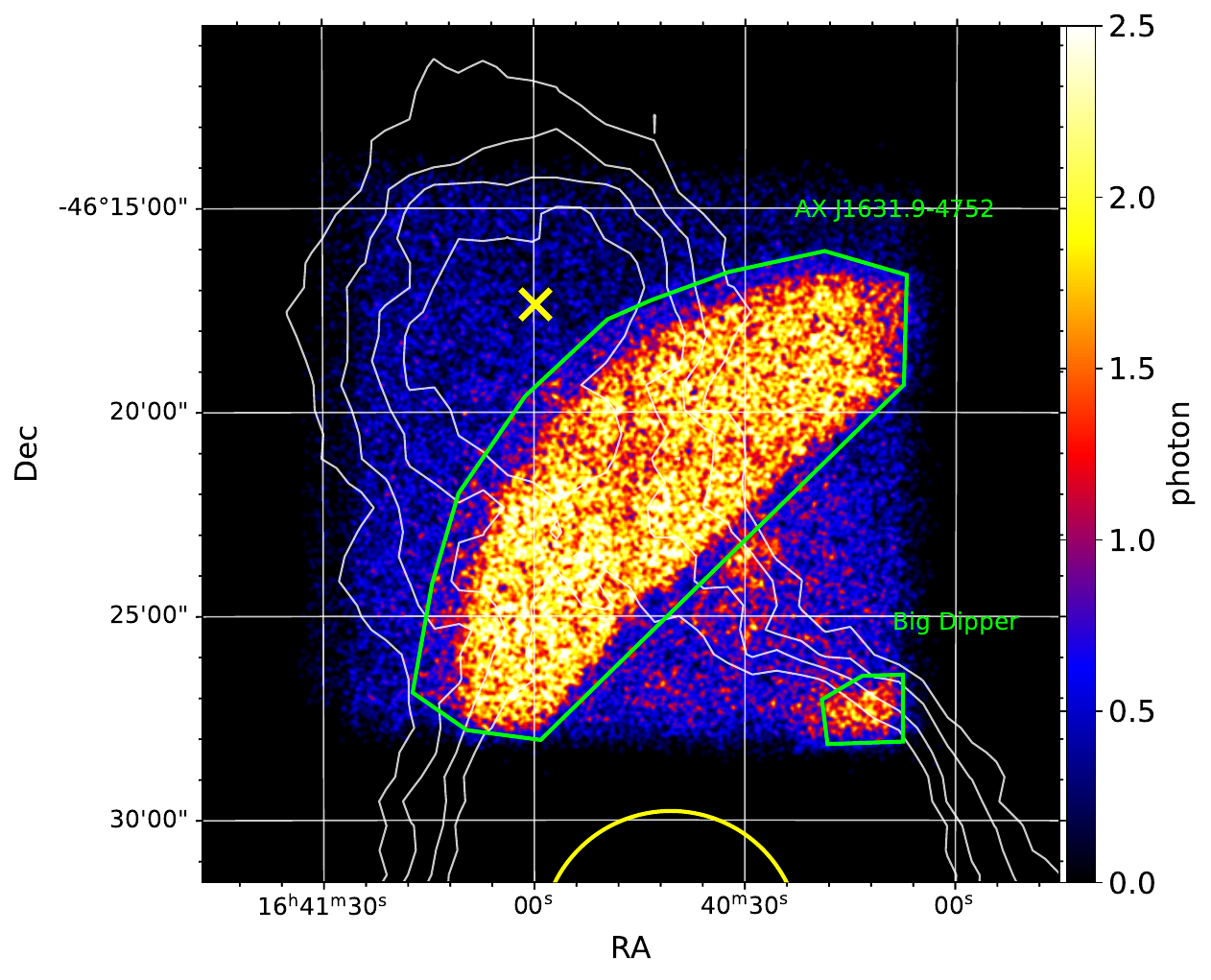}{0.33\textwidth}{(a) FPMA 3--20 keV count map}
    \fig{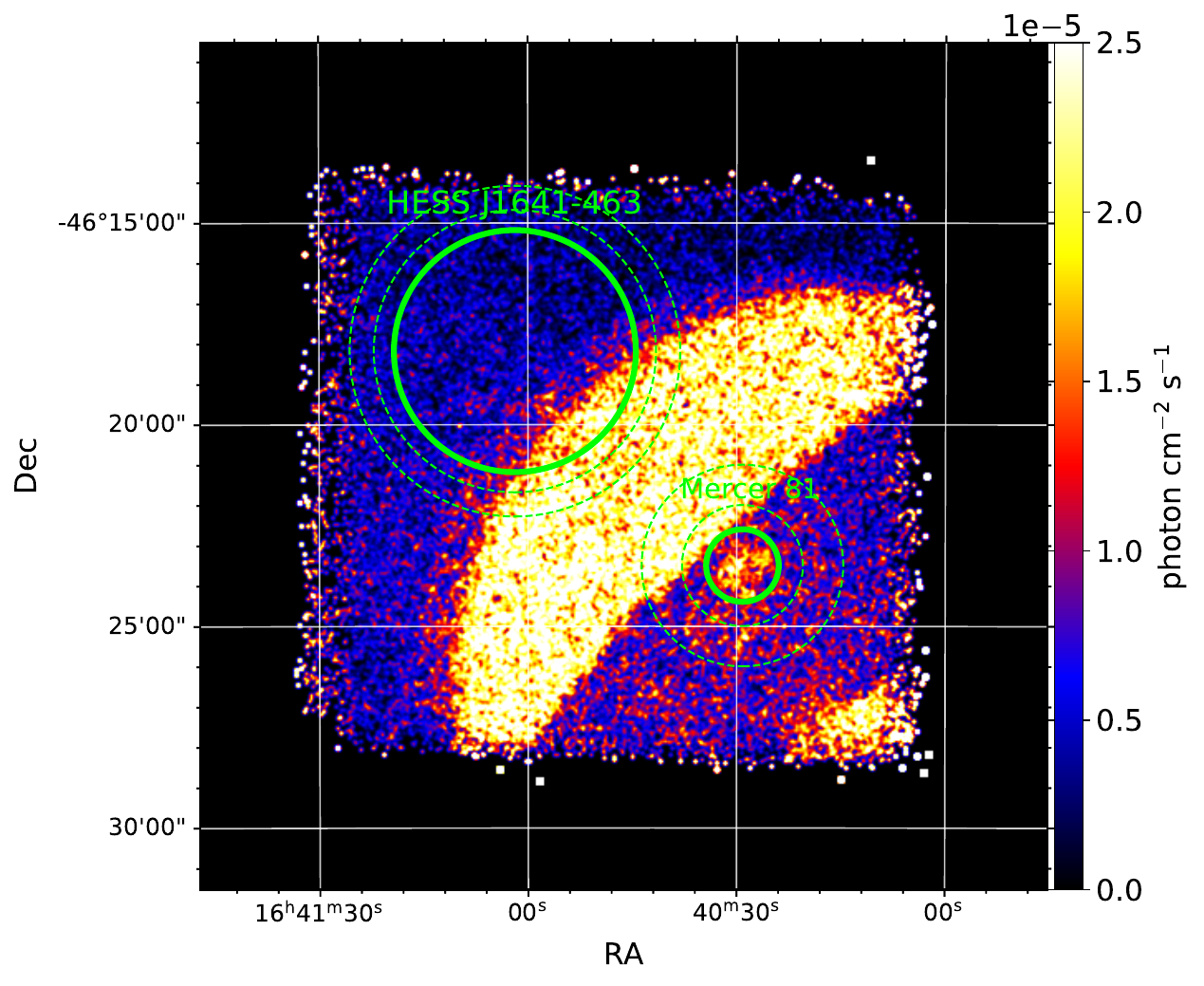}{0.33\textwidth}{(b) FPMA 3--20 keV flux image}
    \fig{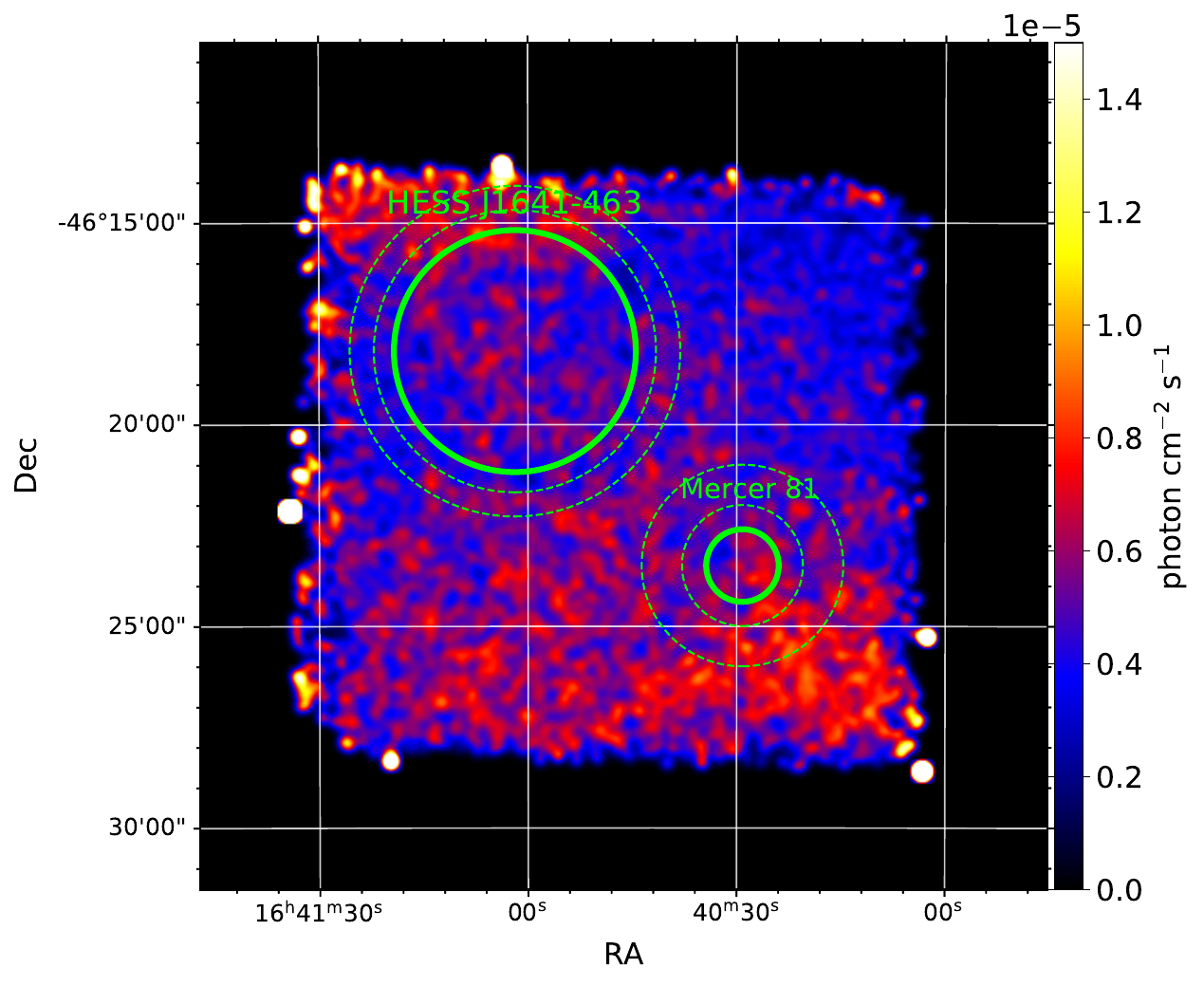}{0.33\textwidth}{(c) FPMA 20--79 keV flux image}
          }
\gridline{
    \fig{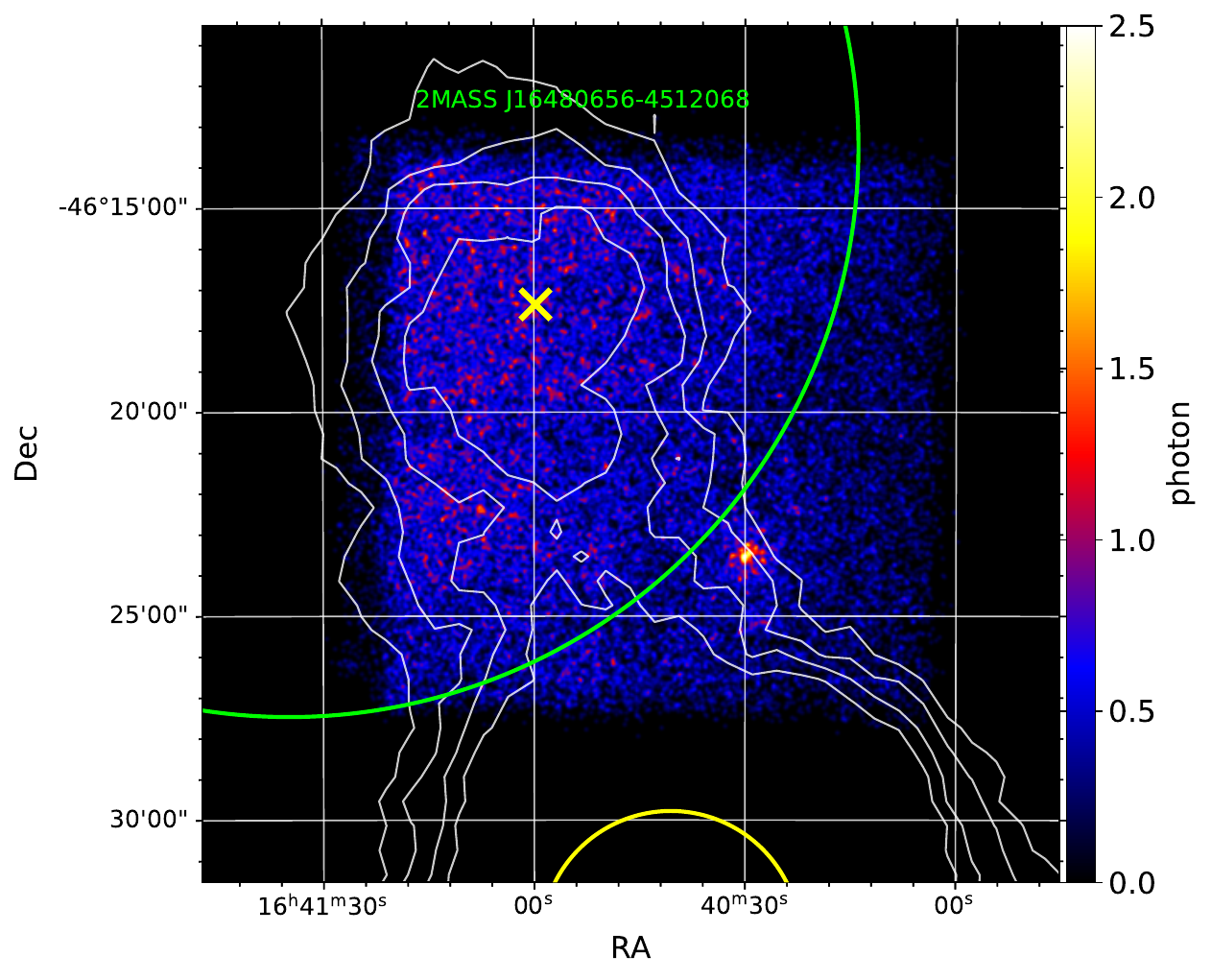}{0.33\textwidth}{(d) FPMB 3--20 keV count map}
    \fig{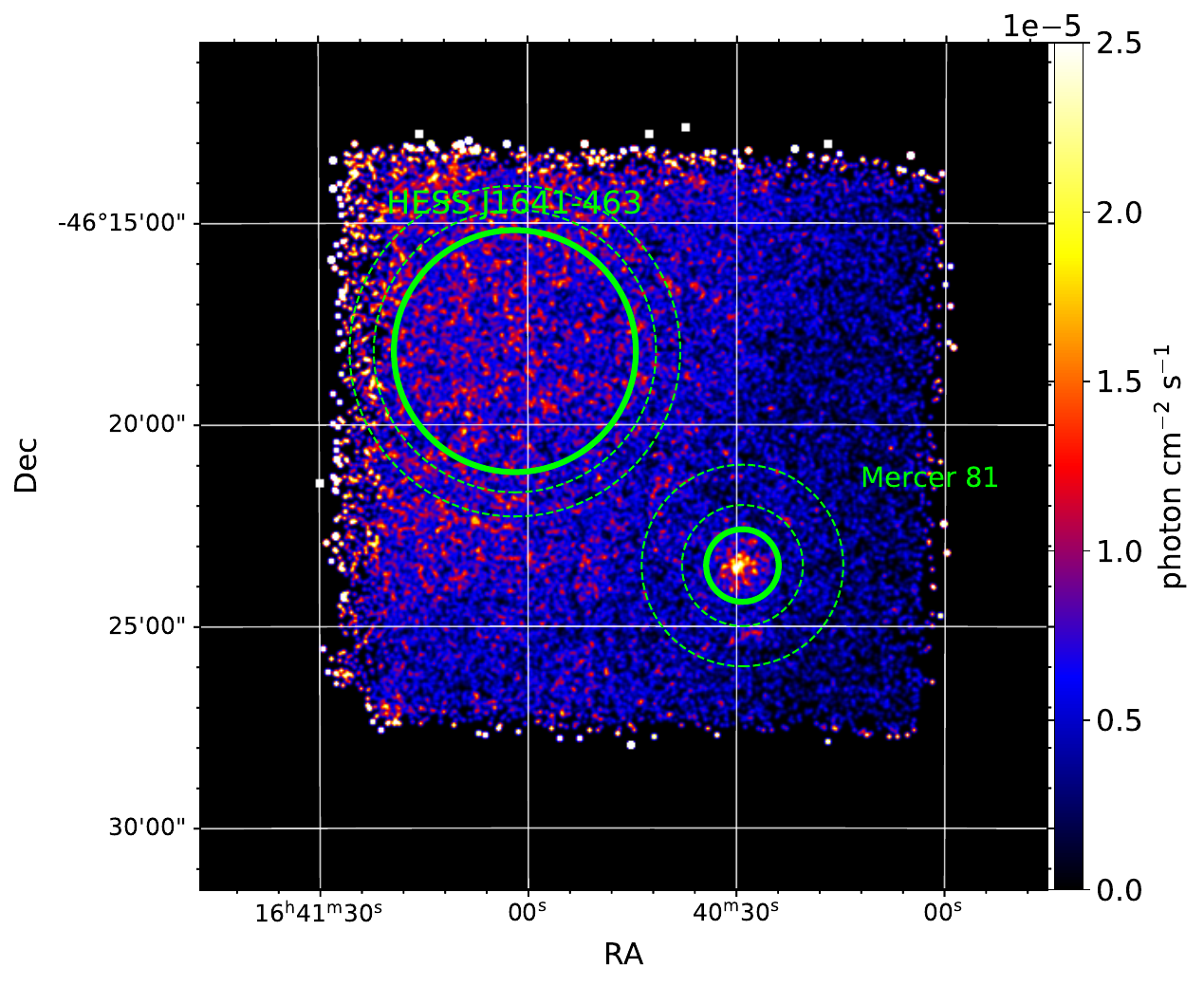}{0.33\textwidth}{(e) FPMB 3--20 keV flux image}
    \fig{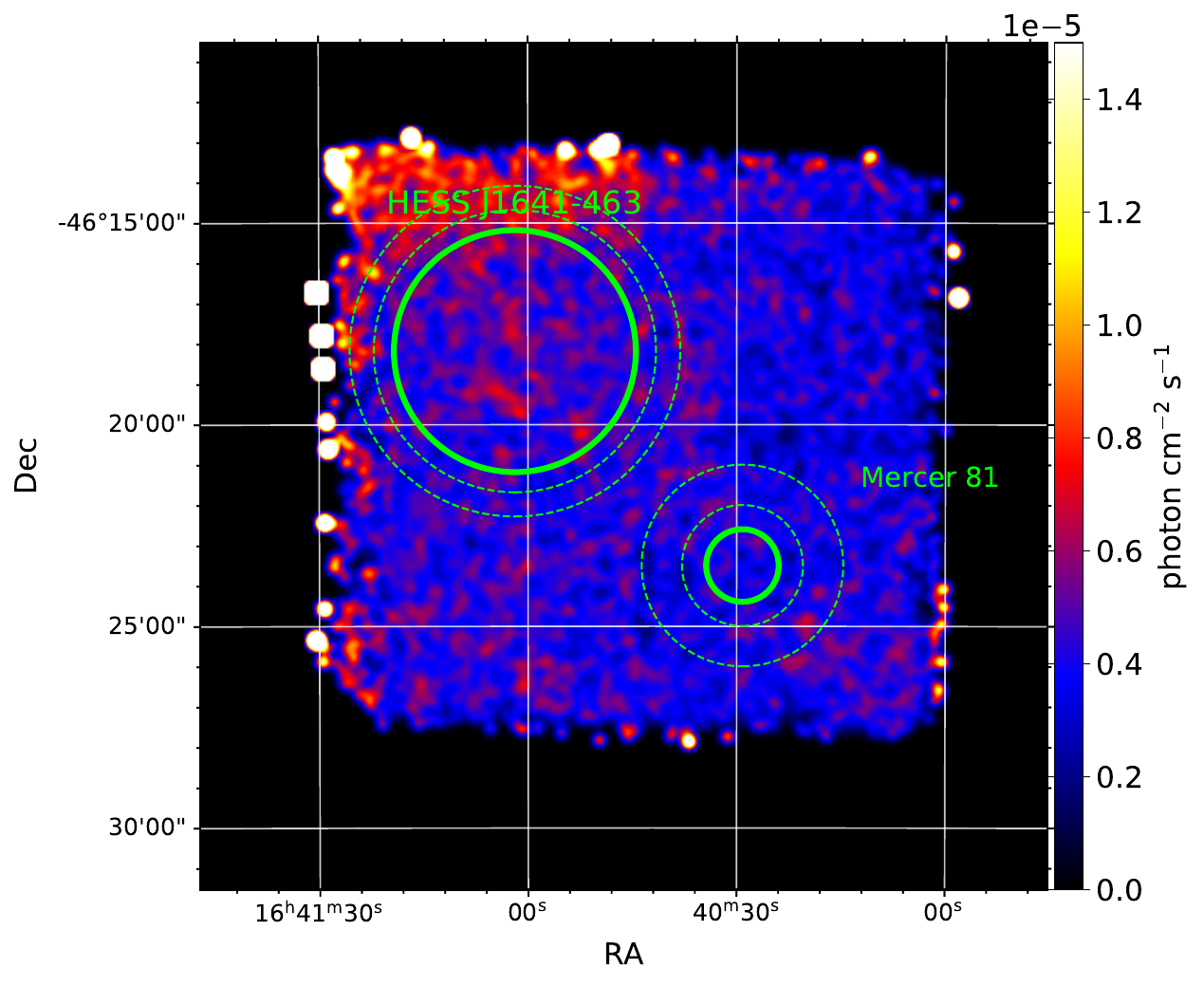}{0.33\textwidth}{(f) FPMB 20--79 keV flux image}
          }
\caption{
Left (a and d): 3--20 keV count maps of \nustar, shown with stray-light sources, the contour of \hess\ significance map \citep{HESS2014_hessj1641}, and a yellow cross of the best-fit position of the GeV gamma-ray source \citep{mares_constraining_2021}.
Middle (b and e): Flux images with \nustar\ in
3--20 keV.
Right (c and f):  Flux images with \nustar\ in
20--79 keV.
The FPMA and FPMB images are respectively shown in the top and bottom panels.
Source and background regions for spectral analysis are indicated by green solid and green dashed lines, respectively.
\label{fig:image_nustar} }
\end{figure}

\figref{fig:image_chandra} shows flux images with \chandra\ in the 0.5--7 keV band, 
and \figref{fig:image_nustar} shows flux images with \nustar\ in the 3--20 and 20--79 keV bands.
There is no apparent counterpart of \j1641 (see Figures \ref{fig:image_chandra} and \ref{fig:image_nustar}).
Another source, stellar cluster candidate Mercer~81, clearly appeared in both the \chandra\ and \nustar\ images.
The \chandra\ images contain the other point-like sources, reported by \cite{HESS2014_hessj1641}.

\nustar\ data are sometimes contaminated by nearby bright sources located outside the \ac{fov}, known as stray-light contamination.
As shown in \figref{fig:image_nustar}, unfortunately our observation was heavily contaminated by stray-light sources, 
since the target region is crowded in the Galactic plane with a low latitude of $b=0\fdg 1$.
The FPMA image included stray light from AX~J1631.9$-$4752 and Big Dipper,
while the FPMB image was contaminated by stray light from 2MASS~J16480656$-$4512068 (see the panels (a) and (d) in \figref{fig:image_nustar}).
Because the TeV gamma-ray emitting region was entirely overwrapped with the stray-light component in the FPMB image, it would be better to use local background in the same \ac{fov} for estimation of the background rather than using Nuskybgd \citep{Wik2014}. 
It should be noted that we checked consistency between the two background methods.
For the spectral analysis in \secref{sec:spectrum}, the stray-light region of AX~J1631.9$-$4752 in the FPMA data was excluded, and we extracted the background in the same region contaminated by stray-light 2MASS~J16480656$-$4512068 in the FPMB data.


\subsection{Spectrum} \label{sec:spectrum}

This section presents the spectral analysis of the \chandra\ and \nustar\ data.
We used {\tt specextract} and {\tt nuproducts} with `extended = yes', respectively, to produce \chandra\ and \nustar\ spectra.
Xspec \cite[][version 12.12.0]{Arnaud1996} was used in spectral fitting.
Errors are shown in 1$\sigma$ confidence level. 
Source and background regions are shown in \figref{fig:image_chandra} and \figref{fig:image_nustar}.

\begin{figure}[ht!]
\centering
\plottwo{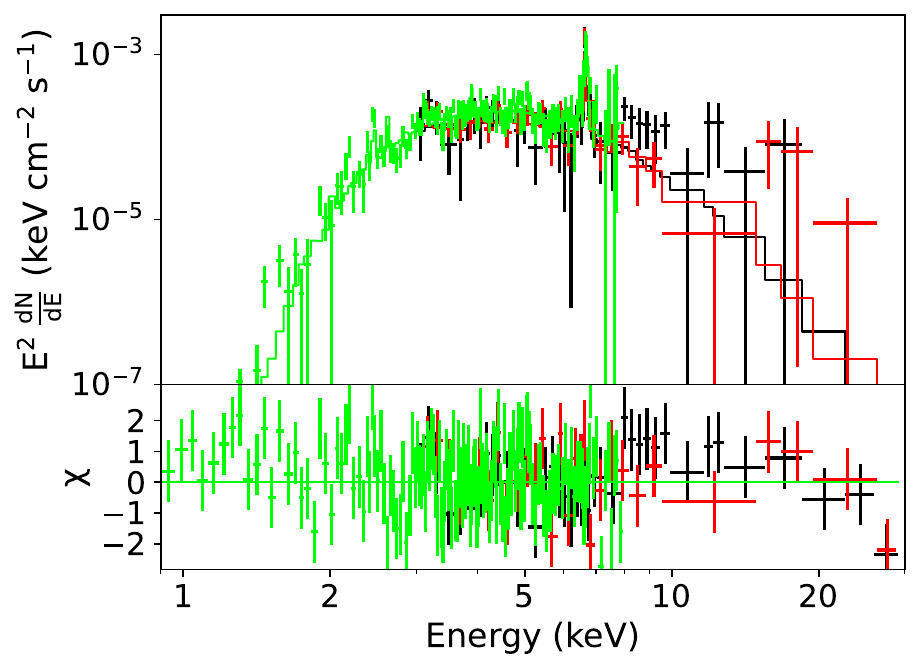}{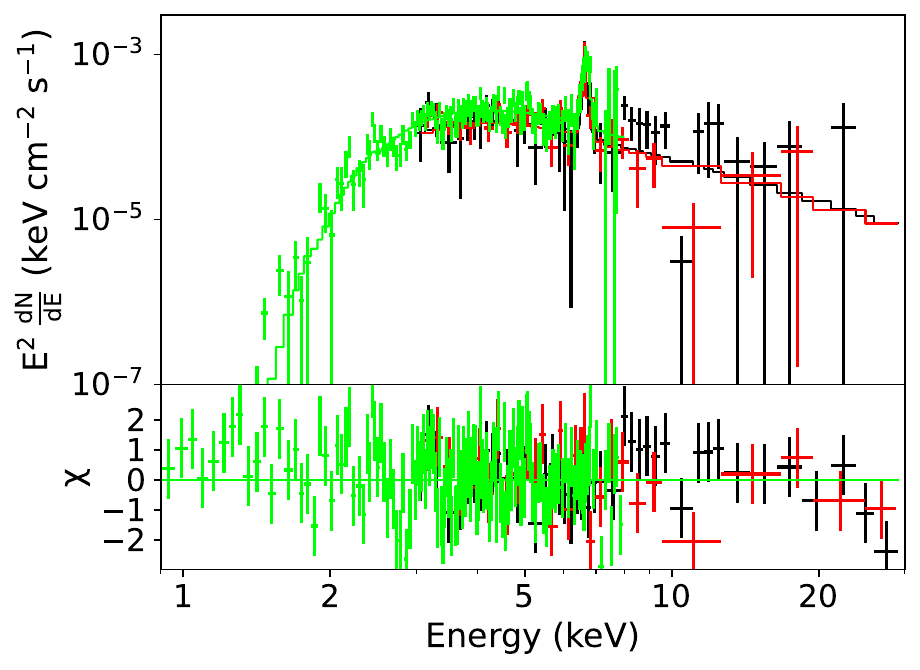}
\caption{
\chandra\ (shown in green) and \nustar\ (FPMA in black and FPMB in red) spectra of Mercer~81, fitted with the thermal model 
(left) and the nonthermal model (right).
The fitting results are shown in \tabref{tab:spectrum_Mercer81}.
\label{fig:mercer81} }
\end{figure}

Besides \j1641, we first conducted spectral analysis of Mercer 81, which was clearly detected by both \chandra\ and \nustar, and thus it is helpful to check consistency between the \chandra\ and \nustar\ analyses.
For the \chandra\ data, the source spectrum was extracted from a circle with a center at (RA, Dec) $=(16\hour40\min29\fs4, -46\degr23\arcmin32\farcs1)$ and a radius of 0\farcm9, while the background region was taken from an annulus with 1\arcmin--2\arcmin.
We used only the data with ObsID 11008, which fully covered the entire emission of Mercer~81. 
For the \nustar\ data, the source region was the same as the \chandra\ data, while the background was extracted from an annulus with 1\farcm5--2\farcm5.
The obtained spectra of Mercer~81 are shown in \figref{fig:mercer81}.
We fit the spectra with two models; nonthermal (i.e., power law plus Gaussian that accounts for $\sim$6 keV line emission) and thermal (apec) models (\tabref{tab:spectrum_Mercer81}).
For both models, absorption was taken into account by using the TBabs model \citep{wilms_absorption_2000}.
When fitting with the nonthermal model, we obtained the best-fit parameters of 
$\NH = (17 \pm 1) \times 10^{22} ~{\rm cm}^{-2}$, 
$\Gamma = 3.9 \pm 0.2 $, and
Gaussian line with center energy of $6.66 \pm 0.03 $ keV, $\sigma = (9.3 \pm 3.7) \times 10^{-2}$ keV, and normalization of $(3.8 \pm 0.6) \times 10^{-6}$ photons~cm$^{-2}$~s$^{-1}$.
The fitting result with the thermal model is as follows:
$\NH = (14 \pm 1) \times 10^{22} ~{\rm cm}^{-2}$, 
$kT = 2.1 \pm 0.2$ keV, and the 
abundance of $1.0 \pm 0.2$.
The thermal model gave a smaller value of $\chi ^2$: 
$\chi ^2$ was 267 (d.o.f. 229) and 258 (d.o.f. 231) for the nonthermal and thermal models, respectively.
We found that the normalization of the \chandra\ and \nustar\ spectra was in good agreement, as shown in \figref{fig:mercer81}, thus our estimation of background (local background method) for the \nustar\ data is reliable to some extent.

\begin{deluxetable*}{ l cc }
\tablecaption{
Fitting results of Mercer 81
\label{tab:spectrum_Mercer81}
}
\tablewidth{0pt}
\tablehead{
\colhead{}  & \colhead{Nonthermal} & \colhead{Thremal} \\
\colhead{Parameter}  & \colhead{(PL+Gauss)} & \colhead{(Apec)} 
}
\startdata
\NH\ (10$^{22}$ cm$^{-2}$) & 17 $\pm$ 1          & 14 $\pm$ 1 \\
$\Gamma$ / $kT$ (keV)   & 3.9 $\pm 0.2$ & 2.1$\pm 0.2$ \\ 
$F_{2-10}$ (10$^{-13}$ \eflux)       & 2.9 $^{+0.2}_{-0.5}$ &  2.9 $^{+0.2}_{-0.4}$ \\ 
Gaussian line energy (keV)   & 6.66 $\pm$ 0.03 & --- \\
Gaussian line width (keV) & $(9.3 \pm 3.7) \times 10^{-2}$ & --- \\
Gaussian normalization (photons~cm$^{-2}$~s$^{-1}$)  & $(3.8 \pm 0.6 ) \times 10^{-6}$ 
  & --- \\
Abundance & --- & 1.0 $\pm$ 0.2 \\
\hline
$\chi^2$ (dof) & 267 (229) & 258 (231) \\
%
\enddata
\end{deluxetable*}

\begin{figure}[ht!]
\centering
\includegraphics[width=0.5\linewidth]{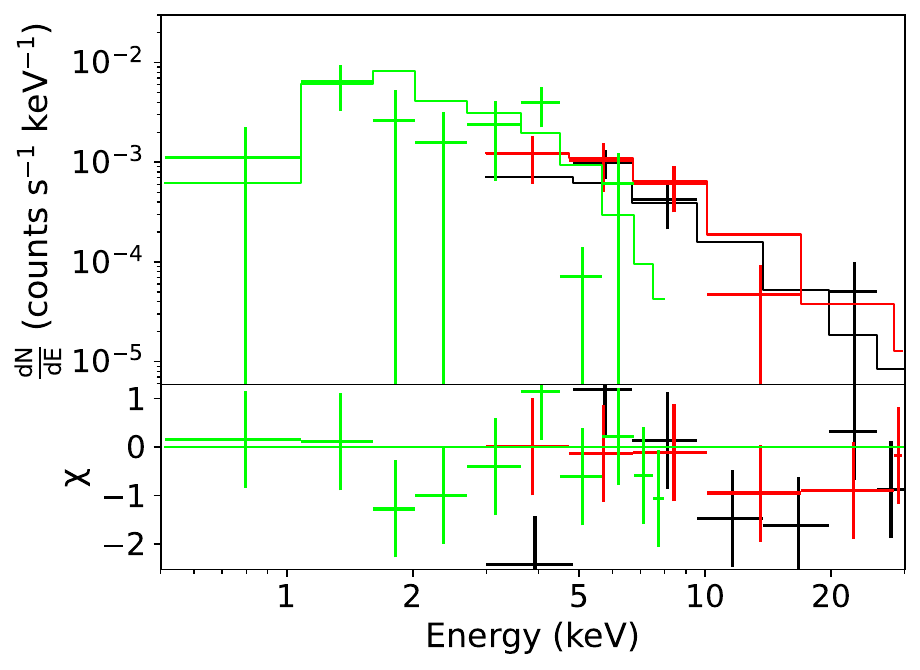}
\caption{
\chandra\ (shown in green) and \nustar\ (FPMA in black and FPMB in red) spectra of \j1641, fitted with the absorbed power-law model.
\label{fig:spec_j1641} }
\end{figure}

The source region of \j1641\ was set to a circle with a center at (RA, Dec) $= (16\hour41\min2\fs1,\ -46\degr18\arcmin13\farcs0)$ and a radius of 3\arcmin, which is comparable to the TeV gamma-ray extent \citep{HESS2014_hessj1641}.
For the \chandra\ data, the background was extracted from an annulus region of 3\farcm3--4\farcm5, centered at the aforementioned position of \j1641. 
For the \nustar\ data, the background was taken from an annulus region of 3\farcm5--4\farcm1 for both FPMA and FPMB.
We show the spectra of \j1641\ in \figref{fig:spec_j1641}.
In the case of a partial coverage, we scaled the obtained normalization to the TeV gamma-ray size of {$r=3$\arcmin} on the assumption of a uniform distribution.
This normalization scaling was applied to the following data:
the \chandra\ data with ObsID 11008, where the region outside the \ac{fov} (46\% of the full coverage) was masked, and
the \nustar\ FPMA data, where the region contaminated by stray-light (36\%) was excluded.

Since there was no apparent X-ray signal from the \j1641\ region, we assumed a fitting model of an absorbed power law and fixed a column density to the Galactic \ion{H}{1} value of $\NH = 2 \times 10^{22}~{\rm cm}^{-2} $ \citep{dickey_h_1990,kalberla_leidenargentinebonn_2005,mares_constraining_2021} and a photon index to $\Gamma = 2$.
On this assumption, 2$\sigma$ upper limits were derived (\tabref{tab:spectrum}).
We checked how the choice of the \NH\ value depends the flux: 
the normalization of the \nustar\ spectrum, which is not largely affected by the absorption, remains within 50\% in a range of \NH\ from $0.7 \times 10^{22}~{\rm cm}^{-2} $  to 20 $ \times 10^{22}~{\rm cm}^{-2}$,
while that of \chandra\ is consistent within a factor of 3 in the same range.
The flux 
with \chandra\ ObsID 12508 in the 2--10 keV energy band was $6.3 \times 10^{-13}$~\eflux, which is reconciled with that with \nustar\ of (6.2--7.1)$ \times 10^{-13}$~\eflux.
The scaled flux obtained with \chandra\ ObsID 11008 is $2.8 \times 10^{-13}$~\eflux, which is in agreement with $2.31 \times 10^{-13}$~\eflux\ in the literature \citep{mares_constraining_2021}.
It should be noted that these values are all consistent within the systematic uncertainty caused by the choice of \NH.

We evaluated X-ray detection significance of \j1641\ as follows.
Background-subtracted count rates are 
$( 1.26 \pm 0.48 ) \times 10^{-2}$ cts s$^{-1}$,
$( 1.87 \pm 1.37 ) \times 10^{-3}$ cts s$^{-1}$, and
$( 6.19 \pm 2.16 ) \times 10^{-3}$ cts s$^{-1}$
for \chandra, \nustar\ FPMA, and FPMB, respectively.
This implies 1.4--2.9$\sigma$ significance above statistical fluctuation.
We also found that $\chi^2$ reduced 
from 501 (d.o.f. 435) to 457 (d.o.f. 434) when fitting the \chandra\ + \nustar\ spectra
with the normalization being fixed to 0 and free, respectively.
This gave us F-test statistic value of 15.7 and probability of $1.38\times 10^{-4}$, corresponding to a 3.5$\sigma$ detection.
Because the detection significance was as low as $\sim 3 \sigma$ and there was no apparent counterpart seen in the images, we conservatively adopt the upper limits in this paper.


\begin{deluxetable*}{ cccccccc }
\tablecaption{
Fitting results of \j1641 
\label{tab:spectrum}
}
\tablewidth{0pt}
\tablehead{
\colhead{Region}  & \colhead{Size} & \colhead{Detector} & \colhead{$F_{2-10}$} & \colhead{$F_{10-20}$}
& \colhead{Scaled $F_{2-10}$}  & \colhead{Scaled $F_{10-20}$} 
}
\startdata
Stray light excluded &	18.1	& \nustar\ FPMA	& 4.0 &	1.7 &	6.2 &	2.7 \\
Full &	 	28.3	& \nustar\ FPMB	& 7.1 & 3.0 & 7.1 & 3.0 \\
Full (ObsID 12508) &   28.3	& \chandra\ ACIS-S  & 5.1 & --- & 6.3 & --- \\
South half (ObsID 11008) &   15.35 & \chandra\ ACIS-I       & 1.5 &--- & 2.8 & --- 
\enddata
\tablecomments{
\\
Size is in units of arcmin$^2$.
\\
Flux is 2$\sigma$ upper limit and in units of $10^{-13}$~\flux. $F_{2-10}$ and $F_{10-20}$ indicate the flux upper limit in the 2--10 keV and 10--20 keV bands, respectively.
\\
`Scaled' means the scaled value to the TeV gamma-ray extension with $r=3\arcmin$.
}
\end{deluxetable*}

\section{Broadband SED modeling} \label{sec:modeling}

\begin{figure}[ht!]
\centering
\includegraphics[width=0.65\linewidth]{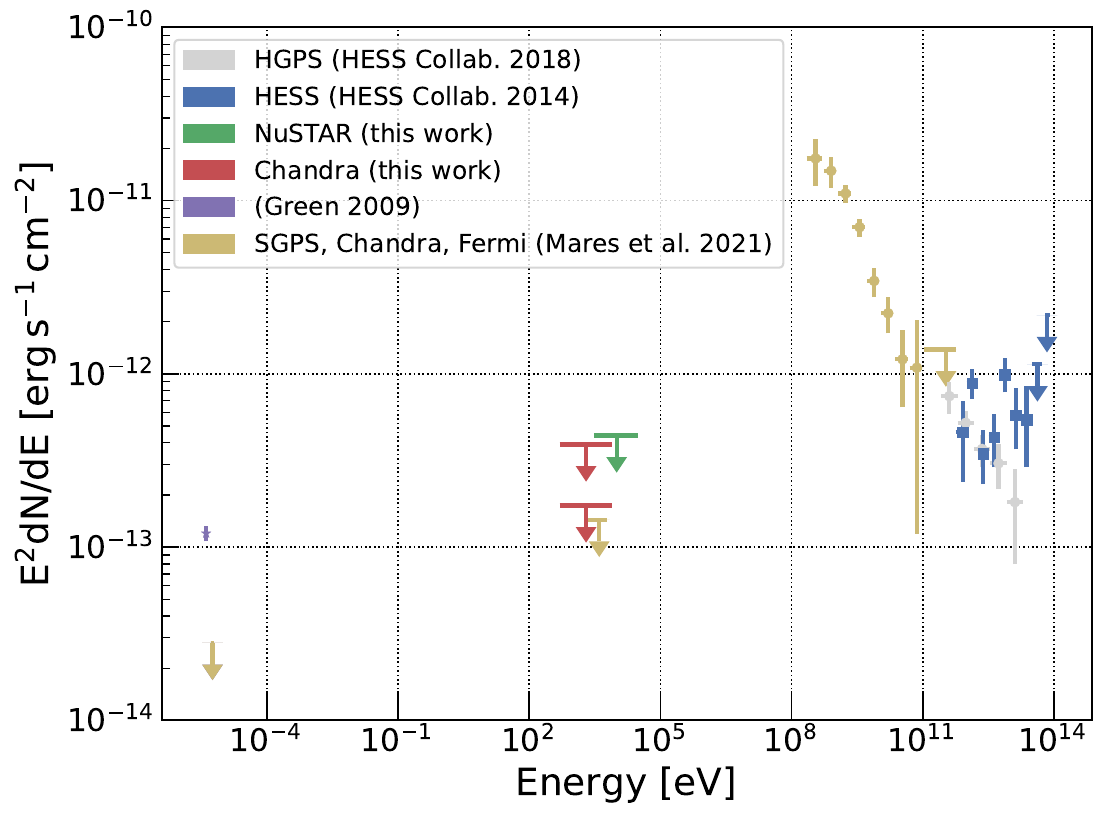}
\caption{
\ac{sed} of \j1641. All data in the literature \citep{HESS2018_HGPS,HESS2014_hessj1641,mares_constraining_2021,Green} are also shown.
\label{fig:sed_all} }
\end{figure}

\figref{fig:sed_all} shows the broadband \ac{sed} of \j1641, including the X-ray upper limits in this paper as well as the radio and gamma-ray flux in the literature.

SNR G338.5$+$0.1, which is a counterpart candidate of \j1641, was identified as a radio SNR by \cite{shaver_galactic_1970} in an investigation of extended radio sources with the Parkes telescope at 5000~MHz and the Molonglo radio telescope at 408~MHz.
However, because of the complex region in the vicinity, the association is uncertain, and the SNR itself has an undefined shape, preventing precise determination of its radio flux and size.
It was reported that G338.5$+$0.1 has a diameter of 5\arcmin--9\arcmin\ and $\sim$12 Jy at 1 GHz \citep{Green,whiteoak_most_1996}.
Although the radio flux of the entire SNR is roughly 12 Jy at 1 GHz, 
\cite{mares_constraining_2021} calculated the radio emission directly associated with the gamma-ray emitting region of \j1641\ (i.e., the size in FWHM of 3\arcmin) based on the Southern Galactic Plane Survey (SGPS) data \citep{haverkorn_southern_2006}, resulting in an upper limit of $\sim$2 Jy at 1.4 GHz.
The former value is used in this paper, otherwise mentioned.
Note that the usage of the radio flux affects only a leptonic modeling, not a hadronic component that is our main point in this paper.

GeV gamma rays are detected with \lat\ in the region of \j1641\ \citep{mares_constraining_2021}.
The GeV gamma-ray source is point-like and has a spectrum with a curvature, implying that it would likely be a pulsar.
Since there is no counterpart of the source at the other wavelength, we do not include the GeV gamma-ray emission in our models.

In the \hess\ Galactic plane survey (HGPS),
the obtained spectrum of \j1641\ was steeper than the one from the analysis dedicated for \j1641\ by \cite{HESS2014_hessj1641}:
the photon index was 2.47$\pm$0.11 \citep{HESS2018_HGPS}.
See \figref{fig:sed_all} for comparison of these two spectra.
In this paper, we mainly use the spectrum of \cite{HESS2014_hessj1641}. 

The upper limits of the X-ray flux obtained in \secref{sec:xray}, combined with the gamma-ray and radio flux in the literature, enable us to model the broadband \ac{sed} (\figref{fig:sed_all}) in order to interpret the origin of the very-high-energy gamma-ray emission.
In the case of a leptonic origin, \ac{ic} scattering of high-energy electrons is accounted for the gamma-ray radiation, and synchrotron radiation from the same electrons produce radio and X-ray photons.
In a hadronic scenario, decays of \pizero\ produced in proton-proton interactions are applied to the gamma-ray data.
Other channels in the proton-proton interactions yield secondary electrons and positrons, which are potentially energetic enough to produce synchrotron radiation in the X-ray energy band.
We summarize derivation of the secondary electrons in \secref{sec:elec} and apply it to the observation in \secref{sec:hadronic}.

\subsection{Secondary electrons in proton-proton interaction} \label{sec:elec}

\begin{figure}[ht!]
\centering
\includegraphics[width=0.65\linewidth]{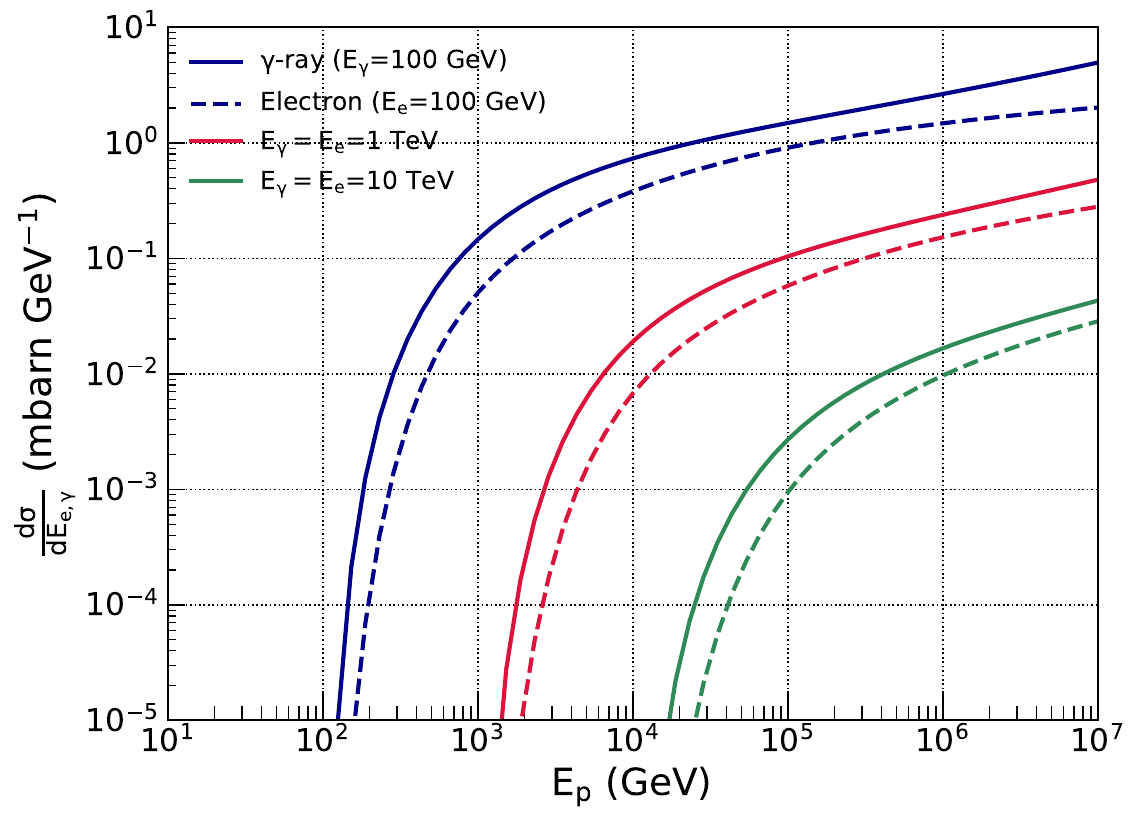}
\caption{
Differential cross section, $\frac{d\sigma}{dE}$, 
of gamma-rays (solid lines)
and secondary \elec\ (dashed lines) 
for $E_e = E_\gamma = 100$ GeV, 1 TeV, and 10 TeV.
\label{fig:dsigma} }
\end{figure}

When \ac{cr} protons collide with protons in ambient medium (proton-proton interaction), pions, \pizero\ and $\pi^\pm$, are produced.
They decay into the following products:
\begin{eqnarray}
\pi^0   & \rightarrow & 2 \gamma  \label{eq:pion_gamma}  \\
\pi^\pm & \rightarrow & \mu^\pm + \nu_\mu (\bar{\nu}_\mu)  \label{eq:2} \\
& & \mu^\pm  \rightarrow  e^\pm + \nu_e (\bar{\nu}_e) + \bar{\nu}_\mu (\nu_\mu ) \label{eq:piton_ele}
\end{eqnarray}
\pizero\ decay yields gamma-ray photons (\eqref{eq:pion_gamma}).
Electrons and positrons (\elec) are also produced from $\pi^\pm$ through Equations \ref{eq:2} and \ref{eq:piton_ele}, and these electrons and positrons are referred to as secondary electrons hereafter.
The secondary electrons have $\sim$10\% energy of the primary protons. 
The cross section of the \pizero-decay photons, $\frac{d\sigma (E_p,E_\gamma)}{dE_\gamma}$, is well formulated in the literature \citep[e.g.,][] {kelner_energy_2006,Kafexhiu2014},
and Geant4-based cross section in \citet{Kafexhiu2014} is adopted for in this paper. 
We make use of a library of LibppGam\footnote{\url{https://github.com/ervinkafex/LibppGam}} for calculating the gamma-ray cross section.
\cite{kelner_energy_2006} derived the spectral ratio of the products of proton-proton interactions, $F_i (E_i / E_p, E_p)$, where the subscript $i$ indicates gamma rays, electrons, or neutrinos 
and $E_p$ is the energy of the primary proton.
The spectrum for the secondary electrons, $F_e  (E_e / E_p, E_p)$, is given by Equation (62) in \cite{kelner_energy_2006}.
Thus, the cross section of the secondary electrons is described by
\begin{eqnarray}
    \frac{d\sigma (E_p,E_e)}{dE_e}  = \frac{d\sigma (E_p,E_\gamma)}{dE_\gamma} \frac{ F_e( \frac{E_e}{E_p}, E_p ) }{ F_\gamma( \frac{E_\gamma}{E_p}, E_p ) }  \label{eq:xsection} .
\end{eqnarray}
\figref{fig:dsigma} compares the cross section of the secondary electrons by \eqref{eq:xsection} with that of the gamma rays in \cite{Kafexhiu2014}.

Following Equation (25) in \cite{Kafexhiu2014},
the energy distribution of the secondary electrons can be expressed as
\begin{eqnarray}
    Q_e (E_e, t) = nc \int dE_p \beta  \epsilon (E_p) \frac{d\sigma (E_p,E_e)}{dE_e} \frac{dN_p}{dE_p}  \label{eq:injection}  .
\end{eqnarray}
Here, $n$ is the ambient density,
$c$ the speed of light, 
and $\frac{dN_p}{dE_p}$ the spectrum of protons.
Technically, we made use of a library of GAMERA\footnote{\url{http://libgamera.github.io/GAMERA/docs/documentation.html}} to solve \eqref{eq:cooling}.
We multiplied the cross section by the so-called nuclear enhancement factor $ \epsilon (E_p)$ to account for proton-nucleus interactions \citep{Kafexhiu2014}.
In this paper, the nuclear enhancement factor was taken from \cite{sihver_total_1993}.

The secondary electrons, injected with \eqref{eq:injection}, suffer radiation cooling.
Therefore, we took the cooling effect into consideration in the following equation of time evolution:
\begin{eqnarray}
    \frac{ \partial N_e(E_e,t) }{ \partial t} = \frac{\partial}{\partial E_e} \left[ b(E_e) N_e(E_e,t) \right] + Q_e(E_e, t),   \label{eq:cooling}
\end{eqnarray}
where $N_e$ is the spectrum of electrons, and $b(E)$ indicates the energy loss rate.
Although $b(E)$ consists of synchrotron radiation, \ac{ic} scattering, Bremsstrahlung, and Coulomb collision, 
energy loss due to synchrotron radiation is dominant for $>$1~GeV electrons for a parameter range we consider in the paper.
In the case of synchrotron radiation, the break energy is derived by equating the radiation timescale to the age of the source ($T$), 
\begin{equation}
    E_b = 125~\mathrm{TeV} ~ \left(\frac{T}{\rm kyr}\right)^{-1}  \left(\frac{B}{10~\mu \mathrm{G} }\right)^{-2} ,
    \label{eq:Eb}
\end{equation}
where $B$ indicates the strength of the magnetic field.

\begin{figure}[ht!]
\plotone{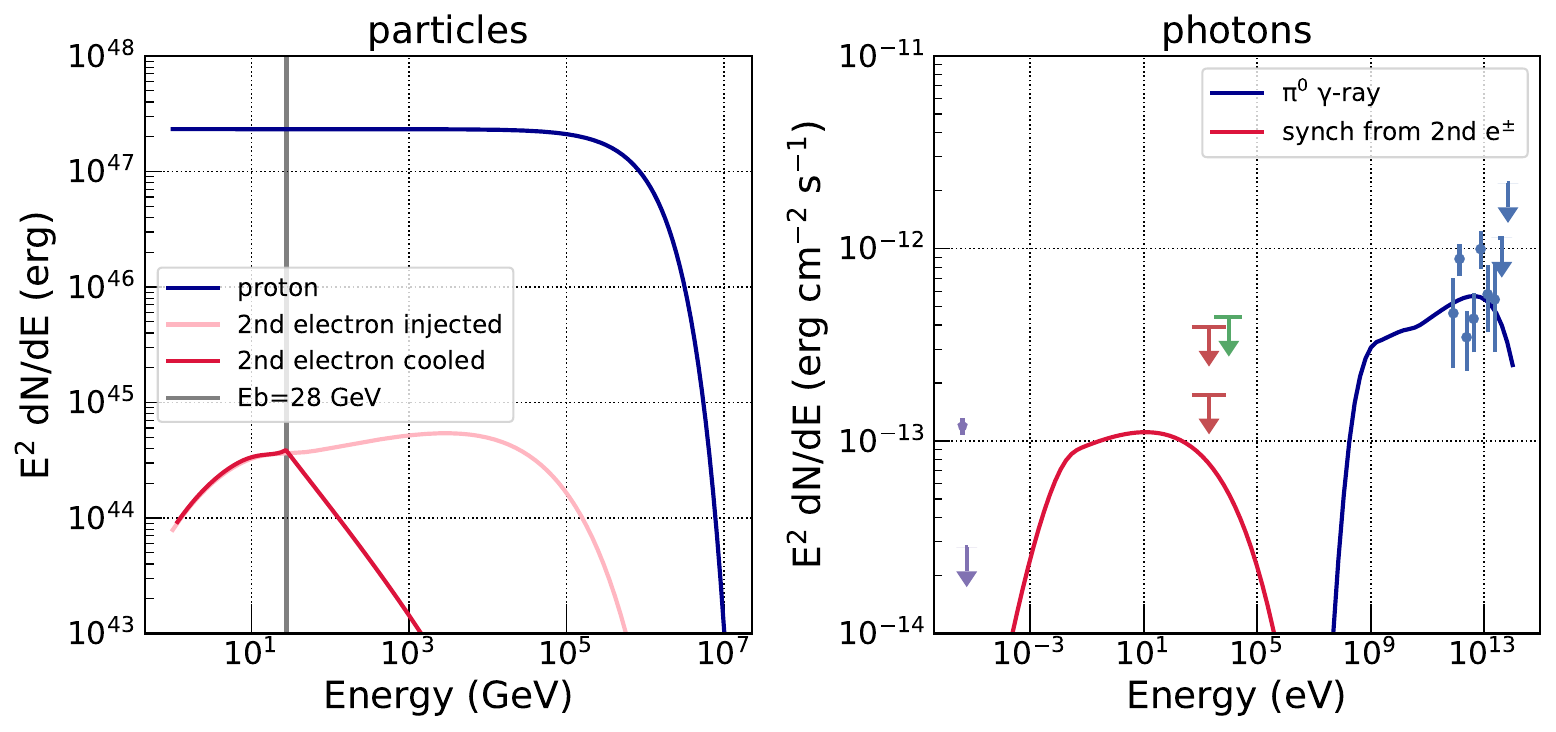}
\caption{
Baseline model of particles (left) and photons (right) with parameters listed in \tabref{tab:baseline}.
Left: the injection spectrum of the protons is shown with the blue line, the injection spectrum of the secondary electrons (\eqref{eq:injection} multiplied by the age $T$) is indicated by the pink line, 
and the spectrum of the cooled secondary electrons is illustrated by the red line.
Right: See \figref{fig:sed_all} for the observational data.
\label{fig:example} }
\end{figure}

\figref{fig:example} shows an example of the model using parameters summarized in \tabref{tab:baseline}.
We assumed a power-law spectrum with an exponential cutoff for primary protons;
\begin{eqnarray}
    \frac{dN_p}{dE_p} = K \left( \frac{E_p}{E_{\rm ref}} \right)^{-s_p} \exp \left[ -\left(\frac{E_p}{E_c}\right) \right] , 
    \label{eq:proton}
\end{eqnarray}
where $K$ is the normalization, 
$E_{\rm ref} $ (=1 TeV) is the reference energy, 
$s_p$ is the spectral index,
and $E_c$ is the cutoff energy.

In \figref{fig:example} (right), we show an example of our radiation models, overlaid with the observation data of the TeV gamma ray \citep{HESS2014_hessj1641}, our X-ray upper limits, and the radio flux \citep{Green,mares_constraining_2021}.
We calculated radiation spectra of \pizero-decay gamma rays from the primary protons and synchrotron X-rays from the secondary electrons
by using Naima\footnote{\url{https://naima.readthedocs.io/en/latest/}} \citep{naima}.
It should be noted that the \ac{ic} component from the secondary electrons is smaller than the pion-decay emission by two orders of magnitude.

\subsection{Application to HESS J1641-463 } \label{sec:hadronic}

\begin{figure}[ht!]
\gridline{
    \fig{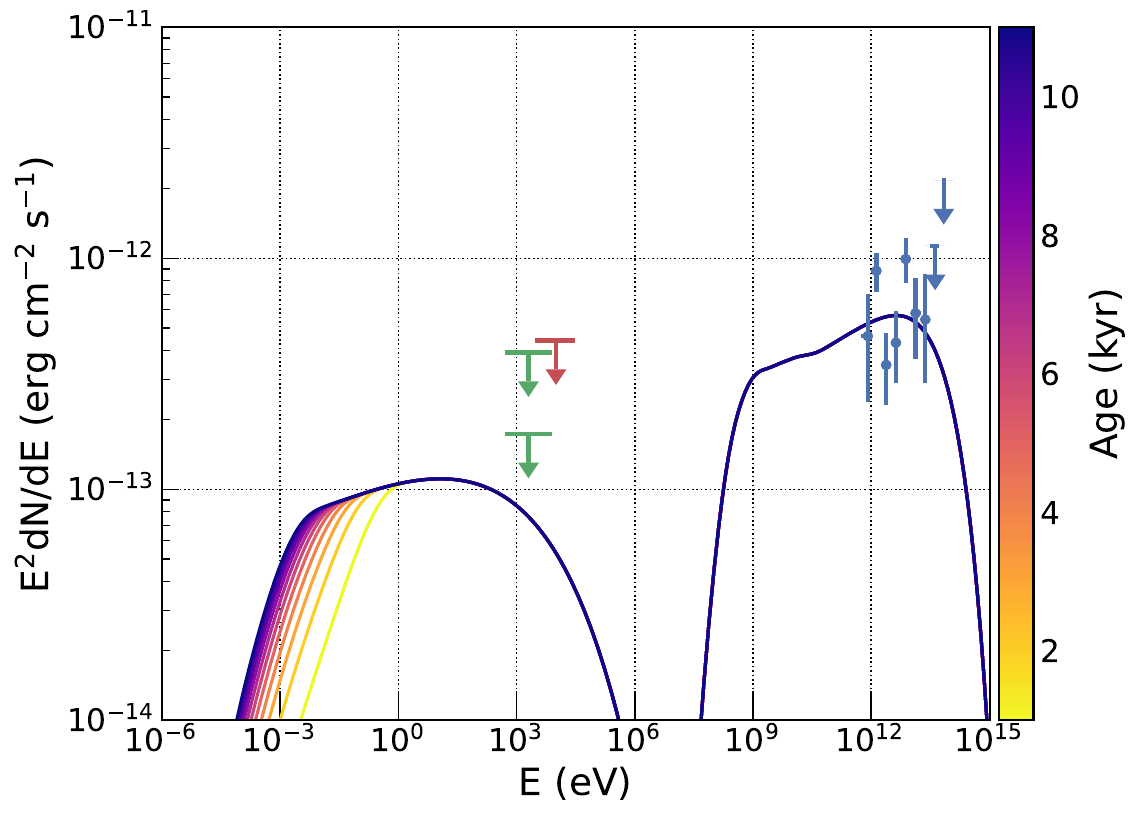}{0.4\textwidth}{(a) Model of \tabref{tab:baseline}, but $T=$1--11 kyr}
    \fig{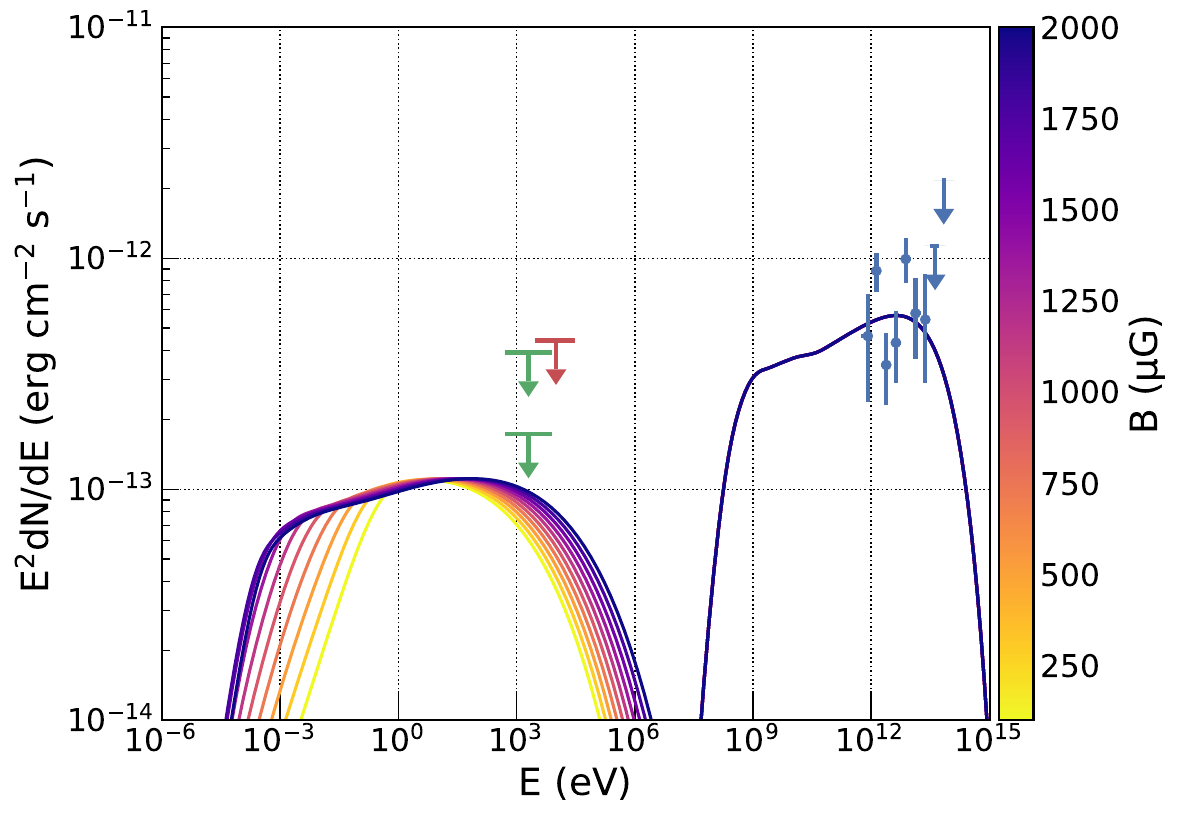}{0.4\textwidth}{(b) Model of \tabref{tab:baseline}, but $B=$100--2000 \uG}
          }
\gridline{
    \fig{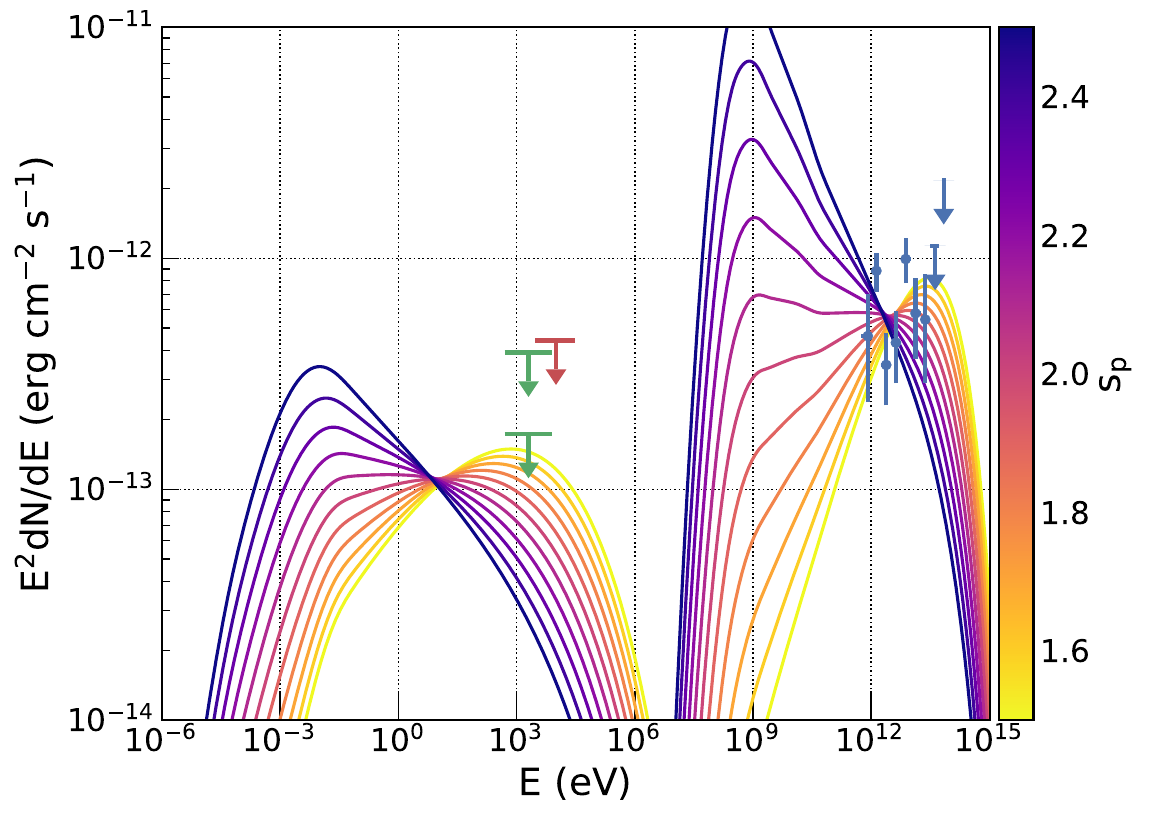}{0.4\textwidth}{(c) Model of \tabref{tab:baseline}, but $s_p=$1.5--2.5. 
    }
    \fig{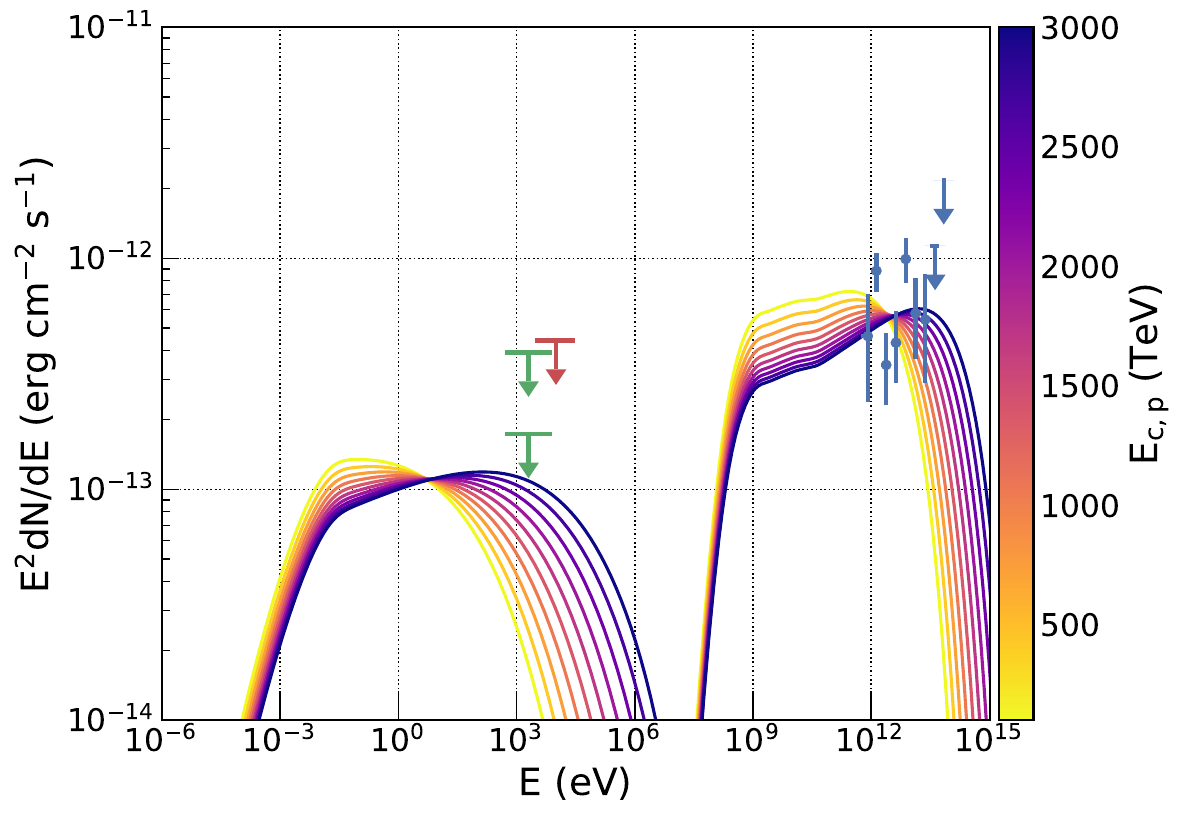}{0.4\textwidth}{(d) Model of \tabref{tab:baseline}, but $E_{c,p}=$100--3000 TeV}
          }
\caption{
Models of \tabref{tab:baseline}, but changing the age $T$ (a), the magnetic field $B$ (b), the spectral index $s_p$ (c), and the maximum energy $E_{c,p}$ (d).
\label{fig:models} }
\end{figure}

%
We adopt  the hadronic model to the \ac{sed} of \j1641.
The hadronic interpretation is supported by the presence of a dense molecular cloud with $n \sim 100$ \cc\ observed by NANTEN \citep{HESS2014_hessj1641}.
Computation of our model requires parameters of $n,~ d$ (distance), $T,~ B,~ s_p,~ E_{c,p}$, and $W_p$.
Although these parameters are poorly constrained, we defined a baseline model as $(n,~ d,~ T,~ B,~ s_p,~ E_{c,p}) = $ (100 \cc, 11 kpc, 5 kyr, 300 \uG, 2.0, 1 PeV), listed in \tabref{tab:baseline}.
From the literature, we adopted the ambient density of 100~\cc, the distance of 11 kpc, and the age of 5 kyr \citep{dickey_h_1990,kalberla_leidenargentinebonn_2005,HESS2014_hessj1641}.
We set $B$ = 300 \uG, assuming that the gamma rays are produced in a molecular cloud where the magnetic field could be from tens to hundreds of \uG\ 
(see \figref{fig:models} for dependency on these values).
$s_p = 2$ and $E_{c,p}$ = 1 PeV are obtained to reproduce the TeV gamma-ray spectrum  \citep{HESS2014_hessj1641}.
First, we fit the TeV gamma-ray data with the \pizero-decay model to determine the normalization of protons, $K$ in \eqref{eq:proton}, which can be transferred to the total energy of protons ($W_p$).
$W_p$ is 1.5$\times 10^{48}$ erg for the baseline model.
Then, we calculated the secondary synchrotron component with the parameters in \tabref{tab:baseline}.
The result of the baseline model is shown in \figref{fig:example}.

In order to explore how the secondary synchrotron emission depends on the parameters,
we performed the aforementioned computations with the parameters different from the baseline model.
We changed $T$ in a range of 1--11 kyr, $B$ in 100--2000 \uG, $s_p$ in 1.5--2.5, and $E_{c,p}$ in 100--3000 TeV.
The results are shown in \figref{fig:models}.
The age, $T$, has an effect only on the break energy of the electrons (i.e., the energy where the age and the time scale of sycnhrotron cooling are balanced in \eqref{eq:Eb}) and does not largely affect the model in the X-ray band (\figref{fig:models}a).
The magnetic field, $B$, slightly modifies the synchrotron model (\figref{fig:models}b).
However, the choice of $B$ does not largely change the normalization of the synchrotron radiation, 
since large $B$ makes the energy loss large and the electron spectrum steep, but, at the same time, makes the synchrotron power high, canceling out the steep spectrum of electrons.
The spectral shape of the primary protons, $s_p$ and $E_{c,p}$, largely change the X-ray spectrum (\figref{fig:models}c and d).
It is worth noting that $n$ and $W_p$ (or $K$) are degenerated, and they do not affect the X-ray spectrum.

\begin{deluxetable}{ lll }
\tablecaption{
Baseline model parameter
\label{tab:baseline}
}
\tablewidth{0pt}
\tablehead{
\colhead{Parameter} & \colhead{} & \colhead{Value}
}
\startdata
Ambient density & $n$ & 100 \cc \\
Distance & $d$ & 11 kpc \\
Age & $T$ & 5 kyr \\
Magnetic field & $B$ & 300 \uG \\
Proton spectral index & $s_p$ & 2.0 \\
Proton cutoff energy & $E_{c,p}$ & 1 PeV \\
Proton total energy & $W_p~ (E>1$ TeV) & 1.5$\times 10^{48}$ erg \\
\enddata
\end{deluxetable}

\section{Discussion} \label{sec:discussion}

\subsection{Gamma-ray origin of HESS J1641-463} \label{sec:discussion1}

\subsubsection{Hadronic scenario}

The hadronic gamma rays trace a molecular cloud (or dense gas) illuminated by cosmic rays.
There are two cases proposed:
(1) SNR-cloud interaction --- particles are accelerated at the SNR shock and interacts with the molecular clouds in the immediate vicinity of the SNR
and (2) CR-cloud interaction --- accelerated particles have escaped the acceleration site, diffusively propagated, and hit nearby molecular clouds.
In the former case, the higher-energy CRs can penetrate into clouds deeply, resulting in the relatively hard CR spectrum \cite[e.g., ][]{Inoue2012,gabici_hadronic_2014,Celli2019,Higurashi2020}.
The SNR shock at such dense environment would produce thermal X-rays.
In the CR-cloud scenario, the spectrum of the escaping particles generally depends on assumptions, such as a time from the escape and a distance from the source and the cloud \citep[e.g., ][]{aharonian_emissivity_1996,gabici_searching_2007,ohira_escape-limited_2010}.
Particularly, a self-consistent calculation for the \j1641/HESS~J1604$-$465 system was conducted in detail by \cite{tang_self-consistent_2015}, showing that the distance affects the low-energy part of the spectrum of the escaping particles.

Since all of the models are lower than the X-ray upper limit obtained in this paper, 
we cannot place tight constraints by the X-ray observations at this time.
As illustrated in \figref{fig:models}, we found that the synchrotron radiation from the secondary electrons is variable in the X-ray energy band, depending strongly on $s_p$ and $E_{c,p}$ and slightly on $B$.
On the other words, a precise measurement of the secondary component can constrain the proton spectrum, which will be possible with the future missions (\secref{sec:future}).

The hard TeV gamma-ray spectrum indicates particle acceleration up to the multi-TeV range, $E_{c,p} >$100 TeV \citep[][and \figref{fig:models}]{HESS2014_hessj1641}.
Given the SNR age of $\sim$5 kyr, the shock velocity should slow down to tens of \kms, making it unlikely to produce such energetic particles at present.
Thus, it implies the acceleration had been achieved in the past in the SNR-cloud scenario.
The CR-cloud scenario, on the other hand, does not require such on-site and present acceleration.
In both cases, these high energy CRs can be hardly confined in the SNR shell and escape the shock first.

Our modeling revealed that the total proton energy, $W_p$ ($E>1$ TeV), is roughly $\sim 10^{48}$ erg on the assumption of $n = 100 $~\cc.
Typically, an SN explosion releases an energy of $E_{\rm SN} \sim 10^{51}$ erg, and a fraction of the energy is converted into acceleration of \acp{cr}
(e.g., $E_{\rm CR} \sim 10^{50}$ erg, assuming $\sim$10\% for the conversion factor).
$W_p \sim 10^{48}$ erg in \j1641\ is smaller than the typical value for SNRs.
In the SNR-cloud scenario, 
this may imply the conversion fraction from $E_{\rm SN}$ to $E_{\rm CR}$ is lower for this source. 
If this SNR is indeed interacting with a molecular cloud, parts of it may be radiative, in which case a lot of energy could have been radiated away already.
Alternatively, the small proton energy might support the escape scenario. 

It should be noted that the lack of bright thermal X-ray emission, which is expected for such SNR with dense gas ($n\sim100$~\cc ), may disfavor the SNR scenario. 
However, the SNR with the age of $\sim$5 kyr might be too old to heat up the gas to keV because of the decelerated shock speed. 
On the other hand, in the escape (CR-cloud) scenario, the accelerated particles have already left the source, which naturally explains the absence of the thermal emission. 
The existence of the thermal emission can be verified by deep observation in the soft X-ray band in the future. 

Here, we note the result when we use the steep spectrum obtained in HGPS \citep{HESS2018_HGPS}.
With $n$ being fixed to 100 \cc, we derived $s_p = 2.44 \pm 0.13$, $E_p >$200 TeV, and $W_p = (1.4 \pm 0.17) \times 10^{48}$ erg.
Using these best-fit parameters of the protons and the others of the baseline model,
the synchrotron radiation from the secondary electrons became about one order of magnitude lower than the model optimized for the harder one in \cite{HESS2014_hessj1641}.

\subsubsection{Other scenarios}

\paragraph{Leptonic scenario} \label{sec:leptonic}

The leptonic-dominated model consists of \ac{ic} scattering and synchrotron radiation from primary \ac{cr} electrons.
Assuming a cutoff power-law model of electrons (\eqref{eq:proton}) and typical values for the photon fields of CMB, NIR, and FIR (i.e., temperatures of 2.72 K, 30 K, and 3000 K, and energy densities of 0.261, 0.5, and 1 eV~cm$^{-3}$, respectively \citep{naima}),
fitting the multiwavelength \ac{sed} results in 
the spectral index ($s_e$) of $2.9 \pm 0.1$, the magnetic field of $2.5^{+0.8}_{-0.5}$ \uG, and the poorly determined cutoff energy of $E_{c,e} > 100$ TeV.
Adopting the upper limit of the radio flux derived from the SGPS data \citep{haverkorn_southern_2006,mares_constraining_2021}, $s_e$ and $E_{c,e}$ cannot be well constrained while $B$ should be less than 2 \uG\ to surpress the synchrotron radiation.

Because the TeV gamma-ray flux is compatible with the X-ray upper limits,
the magnetic field should be less than $\sim$3 \uG\ in the leptonic case. 
This B-field value is comparable to the typical value of the interstellar magnetic field, implying that
no shock compression and/or amplification are required
and that the leptonic case seems very unlikely.
Such a small value of the magnetic field may imply the TeV gamma-ray origin is a pulsar halo. 
The possible presence of the pulsar is also supported from the curved spectrum from the GeV gamma-ray point-like source,
although there are no counterparts of the pulsar at the other wavelengths.
Furthermore, the electron total energy ($W_e$ for $E>$1 TeV) is $(3.6 \pm 0.6) \times 10^{47}$ erg. 
If the electron to proton ratio, \kep, is 0.01, the proton energy would be $\sim 10^{49}$ erg, which is larger then the hadronic case.
For the leptonic component to be dominant, the ambient density should be $<$10 \cc, which conflicts with the observed value of 100 \cc.
In conclusion, the leptonic-dominated case is disfavored, although it is not excluded.

\paragraph{Mixed scenario (one-zone)} \label{sec:mix}

Here, we consider adding primary electrons that have the same spectral model (\eqref{eq:proton}) with the protons, as \textit{one-zone} model.
Radiation cooling of electrons is taken into account.
In the ranges of $E_p =100$--1000 TeV, $B=10$--1000 \uG , and $s = $2--2.5, 
no parameter set was found to be in agreements with all the observations:
(1) \kep\ should be $ < 0.01$ for not exceeding X-ray ULs, but then the synchrotron flux becomes much lower than the radio data,
and (2) if we determine the normalization of the electrons to reproduce the radio flux, the model exceeds the X-ray upper limits or the \lat\ data.
$E_{c,p}$ (proton maximum energy) should be $\gtrsim$100 TeV in order to reproduce the hard spectrum in the TeV gamma-ray energy band, resulting in higher cutoff energy in the spectrum of the primary synchrotron radiation, even though cooling is taken into consideration.
In conclusion, one-zone model is unlikely acceptable.
Adopting the upper limit of the radio flux derived from the SGPS data \citep{haverkorn_southern_2006,mares_constraining_2021} does not largely change the aforementioned result.

\paragraph{Mixed scenario (two-zone)} 

In addition to the protons and the secondary electrons fixed to the baseline model (\tabref{tab:baseline}), we add primary electrons with different spectral shape (i.e., the primary electron index of $s_e \neq s_p$, and the primary electron cutoff energy of $E_{c,e} \neq E_{c,p}$). 
The magnetic field for the primary electrons to emit synchrotron photons (labeled as $B_e$ hereafter) is assumed to be independent of that of the secondary electrons (i.e., $B_e \neq B$).
After the synchrotron radiation from the primary electron is normalized to the radio data,
we searched for parameters ($E_{c,e}$ and $B_e$) with which the total flux of all the radiation components does not exceed the observations of the X-ray upper limits and the GeV emission from the point-like source \citep{mares_constraining_2021}.
Assuming $s_e =2.5$, $B_e$ should be larger than 4 \uG\ to suppress the IC component and $E_{c,e}$ should be less than $\sim$10 TeV not to exceed the X-ray upper limit.
If we adopt for \kep=0.01, $E_{c,e}$ and $B_e$ are constrained to be $\lesssim$4 TeV and 20--40 \uG, respectively.


\subsection{Future prospects}  \label{sec:future}

In the hadronic scenario, our models (\figref{fig:models}) suggest that the flux of the secondary synchrotron component is $> 1.8 \times 10^{-14}$ \flux\ in 2--10 keV and $> 4.9 \times 10^{-15}$ \flux\ in 10--50 keV for $E_{c,p} > 100$ TeV.
This flux level is sufficiently detectable with X-ray telescopes.
In order to avoid a contamination of the primary synchrotron radiation which generally has a cutoff at the soft X-ray band, 
hard X-ray detectors would be needed.
Proposed missions, such as HEX-P\footnote{\url{https://hexp.org/}} \citep{madsen_hex-p_2019,madsen_high_2023,mori_high_2023,reynolds_high_2023} 
or FORCE \citep{mori_broadband_2016} 
will be able to measure the emission and determine the proton maximum energy by the X-ray observation for the first time.
Long-exposure observations with the existing X-ray telescopes would also be helpful to search for the secondary synchrotoron radiation, 
and we are planning an \xmm\ observation.


Besides the synchrotron emission from the secondary electrons,
neutrino is also a smoking gun to evidence hadronic acceleration. 
Although the neutrino amount from this single source would be much smaller than the detection threshold of the existing neutrino observatories such as IceCube,
neutrinos accumulated by the hadronic PeVatron candidates may have a contribution to the neutrino diffuse flux detected from the Galactic plane \citep{icecube_collaboration_observation_2023}
as unresolved sources.

\section{Conclusion} \label{sec:conclusion}
We analyzed newly taken \nustar\ data and the archival \chandra\ data of the unidentified TeV gamma-ray source, \j1641,
to search for the synchrotron radiation from the secondary electrons in the hadronic interpretation. 
No X-ray counterpart was detected in the gamma-ray extension, and we obtained 2$\sigma$ upper limits of 
$\sim 6\times 10^{-13}$~\flux\ in the 2--10 keV band and $\sim 3\times 10^{-13}$~\flux\ in the 10--20 keV band.
Combined with the published radio and gamma-ray spectra, 
we performed modeling of the broadband \ac{sed}.
Although the X-ray upper limit could not constrain the hadronic model, we found the flux of the secondary synchrotron component is larger than $\sim 10^{-14}$ \flux, which would be detectable with 
a deep X-ray observation with existing telescopes or with future mission.


\begin{acknowledgments}
We thank the anonymous referee for the helpful comments.
We are greatful to Kaya Mori for the fruitful discussion
and Hiroya Yamaguchi, Hiroyasu Tajima, and Benson Guest for the preparation of the X-ray proposals.
This work has made use of data from the \nustar\ mission, a project led by the California Institute of Technology, managed by the Jet Propulsion Laboratory, and funded by the National Aeronautics and Space Administration.
This work is supported by the NuSTAR Cycle 4 observation program. We appreciate the NuSTAR Operations, Software, and Calibration teams for support with the execution and analysis of these observations.
This paper employs a list of Chandra datasets, obtained by the Chandra X-ray Observatory, contained in~\dataset[DOI:10.25574/cdc.217]{https://doi.org/10.25574/cdc.217}.
This work was supported by Japan Society for the Promotion of Science (JSPS) KAKENHI Grant Numbers JP22K14064 (N.T.), JP19H01936, and JP21H04493 (T.T.).
S.S.H. acknowledges support from the Natural Sciences and Engineering Research Council of Canada (NSERC) through the Canada Research Chairs and the Discovery Grants programs, from the Canadian Institute for Theoretical Astrophysics and from the Canadian Space Agency.

\end{acknowledgments}

%

\vspace{5mm}

\facilities{
\chandra, \nustar
}

\software{
HEAsoft (v6.29), 
CIAO (v4.14, \cite{fruscione_ciao_2006}),
NuSTARDAS (v2.1.1), 
nuskybgd (\url{https://github.com/NuSTAR/nuskybgd} and \url{https://github.com/achronal/nuskybgd-py}; \cite{Wik2014}), 
XSPEC (v12.12.0, \cite{Arnaud1996}),
naima (\url{https://naima.readthedocs.io/en/latest/}; \cite{naima}), 
GAMERA (\url{http://libgamera.github.io/GAMERA/docs/documentation.html}),
}






\bibliography{references}



\end{document}

%% file: mypreset.tex
\def\figref#1{Figure~\ref{#1}} 
\def\tabref#1{Table~\ref{#1}} 
\def\eqref#1{Equation~\ref{#1}} 
\def\secref#1{Section~\ref{#1}} 

\def\comment#1{\textcolor{cyan}{\bf [NT] #1}} %
\def\revise#1{\textcolor{red}{\bf #1}}

\def\j1641{HESS~J1641$-$463}

\def\chandra{{\it Chandra}}
\def\nustar{NuSTAR}
\def\xmm{{\it XMM-Newton}}

\def\lat{{\it Fermi}-LAT}
\def\hess{H.E.S.S.}

\def\elec{\hbox{$e^\pm$}}

\def\uG{\hbox{$\mu \mathrm{G}$}}
\def\flux{\hbox{${\rm erg}~{\rm cm}^{-2}~{\rm s}^{-1}$}}
\def\eflux{\hbox{${\rm erg}~{\rm cm}^{-2}~{\rm s}^{-1}$}} 

\def\kms{\hbox{${\rm km}~ {\rm s}^{-1}$}}
\def\cc{\hbox{${\rm cm}^{-3}$}}

\def\NH{\hbox{$N_{\rm H}$}}

\def\pizero{\hbox{$\pi^0$}}

\def\kep{$K_{\rm ep}$}

\usepackage{acro}

\DeclareAcronym{gde}{
  short = GDE ,
  long  = Galactic Diffuse Emission,
}

\DeclareAcronym{cgb}{
  short = CGB ,
  long  = cosmic gamma-ray background,
}

\DeclareAcronym{grb}{
  short = GRB ,
  long  = gamma-ray burst,
}
 
\DeclareAcronym{agn}{
  short = AGN ,
  long  = active galactic nucleus ,
  long-plural-form = active galactic nuclei
}
\DeclareAcronym{fsrq}{
  short = FSRQ ,
  long  = flat-spectrum radio quasar,
}
  
\DeclareAcronym{lmxb}{
  short = LMXB ,
  long  = low mass X-ray binary ,
  long-plural-form = low mass X-ray binaries ,
}
\DeclareAcronym{hmxb}{
  short = HMXB ,
  long  = high mass X-ray binary ,
  long-plural-form = high mass X-ray binaries ,
}
  
\DeclareAcronym{ic}{
  short = IC ,
  long  = inverse Compton ,
}

\DeclareAcronym{nustar}{
  short = {\it NuSTAR} ,
  long  = Nuclear Spectroscopic Telescope Array ,
}

\DeclareAcronym{bi}{
  short = BI ,
  long  = backside illumination ,
}
\DeclareAcronym{fi}{
  short = FI ,
  long  = frontside illumination ,
}
\DeclareAcronym{fov}{
  short = FoV ,
  long  = field of view ,
}
  
\DeclareAcronym{sim}{
  short = SIM ,
  long  = Science Instrument Module ,
}
\DeclareAcronym{hetg}{
  short = HETG ,
  long  = High Energy Transmission Grating ,
}
\DeclareAcronym{letg}{
  short = LETG ,
  long  = Low Energy Transmission Grating ,
}
\DeclareAcronym{hrc}{
  short = HRC ,
  long  = High Resolution Camera ,
}
\DeclareAcronym{acis}{
  short = ACIS ,
  long  = Advanced CCD Imaging Spectrometer ,
}
\DeclareAcronym{hrma}{
  short = HRMA ,
  long  = High Resolution Mirror Assembly,
}

\DeclareAcronym{compton}{
  short = Compton ,
  long  = Compton Gamma Ray Observatory,
}
\DeclareAcronym{hst}{
  short = HST ,
  long  = Hubble Space Telescope ,
}
\DeclareAcronym{iact}{
  short = IACT ,
  long  = Imaging Atmospheric Cherenkov Telescope ,
}
\DeclareAcronym{wcd}{
  short = WCD ,
  long  = Water Cherenkov Detector ,
}

\DeclareAcronym{hawc}{
  short = HAWC ,
  long  = High-Altitude Water Cherenkov ,
}
\DeclareAcronym{cta}{
  short = CTA ,
  long  = Cherenkov Telescope Array ,
}

\DeclareAcronym{em}{
  short = EM ,
  long  = electromagnetic ,
}
\DeclareAcronym{ism}{
  short = ISM ,
  long  = interstellar medium ,
}
\DeclareAcronym{csm}{
  short = CSM ,
  long  = circumstellar medium ,
}
\DeclareAcronym{sne}{
  short = SNe ,
  long  = supernovae , 
}

\DeclareAcronym{iss}{
  short = ISS ,
  long  = International Space Station ,
}

\DeclareAcronym{uhecr}{
  short = UHECRs ,
  long  = ultra high energy cosmic rays , 
}
\DeclareAcronym{ta}{
  short = TA ,
  long  = Telescope Array , 
}
\DeclareAcronym{auger}{
  short = Auger ,
  long  = Pierre Auger Observatory , 
}
\DeclareAcronym{ams}{
  short = AMS ,
  long  = Alpha Magnetic Spectrometer , 
}
\DeclareAcronym{pamela}{
  short = PAMELA ,
  long  = Payload for Antimatter Matter Exploration and Light-nuclei Astrophysics , 
}
   
\DeclareAcronym{cmb}{
  short = CMB ,
  long  = Cosmic Microwave Background , 
}
\DeclareAcronym{sed}{
  short = SED ,
  long  = spectral energy distribution , 
}
\DeclareAcronym{mhd}{
  short = MHD ,
  long  = magnetohydrodynamical ,
}
\DeclareAcronym{dof}{
  short = dof ,
  long  = degree of freedom ,
}
\DeclareAcronym{cco}{
  short = CCO ,
  long  = central compact object ,
  first-style = default
}
\DeclareAcronym{lmc}{
  short = LMC ,
  long  = Large Magellanic Cloud ,
}
\DeclareAcronym{smc}{
  short = SMC ,
  long  = Small Magellanic Cloud ,
}

\DeclareAcronym{hess}{
  short = H.E.S.S. ,
  long  = High Energy Stereoscopic System ,
  first-style = default
}
\DeclareAcronym{snr}{
  short = SNR ,
  long  = supernova remnant ,
}
\DeclareAcronym{pwn}{
  short = PWN ,
  short-plural = e ,
  long  = pulsar wind nebula ,
  long-plural  = e ,
}

\DeclareAcronym{sn}{
  short = SN ,
  short-plural = e ,
  long  = supernova ,
  long-plural  = e ,
  first-style = default
}

\DeclareAcronym{nw}{
  short = NW ,
  long  = northwest ,
  first-style = default
}
\DeclareAcronym{hxc}{
  short = HXC ,
  long  = hard X-ray component ,
  first-style = default
}

\DeclareAcronym{cr}{
  short = CR ,
  long  = cosmic ray ,
}
\DeclareAcronym{psf}{
  short = PSF ,
  long  = point spread function ,
}
\DeclareAcronym{hpd}{
  short = HPD ,
  long  = half power diameter ,
}
\DeclareAcronym{fwhm}{
  short = FWHM ,
  long  = full width of half maximum ,
}

\DeclareAcronym{pic}{
  short = PIC ,
  long  = particle-in-cell ,
  tag = numerical ,
}
\DeclareAcronym{cxb}{
  short = CXB ,
  long  = Cosmic X-ray Background ,
}
\DeclareAcronym{grxe}{
  short = GRXE ,
  long  = Galactic Ridge X-ray Emission ,
}
\DeclareAcronym{pa}{
  short = PA ,
  long  = Positional Angle ,
}
\DeclareAcronym{dsa}{
  short = DSA ,
  long  = diffusive shock acceleration ,
}

\def\NH{\hbox{$N_{\rm H}$}}

\def\pizero{\hbox{$\pi^0$}}

\def\degr{\hbox{$^\circ$}}
\def\arcmin{\hbox{$^\prime$}}

\def\utw{\smash{\rlap{\lower5pt\hbox{$\sim$}}}}
\def\udtw{\smash{\rlap{\lower6pt\hbox{$\approx$}}}}

\def\fs{\hbox{$.\!\!^{\rm s}$}}
\def\hour{\hbox{$^{\rm h}$}}
\def\min{\hbox{$^{\rm m}$}}
\def\fdg{\hbox{$.\!\!^\circ$}}

\def\farcm{\hbox{$.\mkern-4mu^\prime$}}
\def\farcs{\hbox{$.\!\!^{\prime\prime}$}}

\def\apjl{ApJ}%
\def\apjs{ApJS}%
\def\ao{Appl.~Opt.}%
\def\aap{A\&A}%
